\documentclass[aps,prb,floatfix,twocolumn,superscriptaddress,showpacs,amsmath,amssymb]{revtex4}
\usepackage{graphicx}
\usepackage{epsfig} 
\allowdisplaybreaks

\newcommand{\be}{\begin{equation}}
\newcommand{\ee}{\end{equation}}
\newcommand{\bea}{\begin{eqnarray}}
\newcommand{\eea}{\end{eqnarray}}
\newcommand{\ba}{\begin{eqnarray*}}
\newcommand{\ea}{\end{eqnarray*}}
\newcommand{\dagga}{{\phantom{\dagger}}}
\newcommand{\bR}{\mathbf{R}}

\newcommand{\bq}{\mathbf{q}}

\newcommand{\bRp}{\mathbf{R'}}

\newcommand{\dis}{\displaystyle}

\newcommand{\up}{\uparrow}
\newcommand{\down}{\downarrow}
\newcommand{\fract}[2]{\frac{\dis #1}{\dis #2}}
\newcommand{\Tr}{\mathrm{Tr}}
\newcommand{\eqn}[1]{(\ref{#1})}
\newcommand{\m}[1]{\mathcal{#1}} 
\newcommand{\eps}{\varepsilon}

\begin{document}

\title{Quantum Quenches in the Hubbard Model: Time Dependent Mean Field Theory and The Role of Quantum Fluctuations} 
\author{Marco Schir\'o} \email{mschiro@princeton.edu}

\affiliation{Princeton
  Center for Theoretical Science and Department of Physics, Joseph Henry
  Laboratories, Princeton University, Princeton, NJ
  08544} 
\affiliation{International
  School for Advanced Studies (SISSA), and CRS Democritos, CNR-INFM,
  Via Beirut 2-4, I-34014 Trieste, Italy} 

\author{Michele Fabrizio} 
\affiliation{International School for
  Advanced Studies (SISSA), and CRS Democritos, CNR-INFM, Via Beirut
  2-4, I-34014 Trieste, Italy} 
\affiliation{The Abdus Salam
  International Centre for Theoretical Physics (ICTP), P.O.Box 586,
  I-34014 Trieste, Italy}

\date{\today} 


\begin{abstract}
We study the non equilibrium dynamics in the fermionic 
Hubbard model after a sudden change of the interaction strength.
To this scope, we introduce a time dependent variational approach 
in the spirit of the Gutzwiller ansatz. At the saddle-point approximation, 
we find at half filling a sharp transition between two 
different regimes of small and large coherent oscillations, 
separated by a critical line of quenches where the system is found to relax. 
Any finite doping washes out the transition, leaving aside just a sharp 
crossover. 
In order to investigate the role of quantum fluctuations, 
we map the model onto an auxiliary Quantum Ising Model in a transverse field 
coupled to free fermionic quasiparticles. Remarkably, the Gutzwiller 
approximation turns out to correspond to the mean field decoupling of this 
model in the limit of infinite coordination lattices.  
The advantage is that we can go beyond mean field and include gaussian 
fluctuations around the non equilibrium mean field dynamics. 
Unlike at equilibrium, we find that quantum fluctuations 
become massless and eventually unstable before the mean field dynamical critical line, which 
suggests they could even alter qualitatively the mean field scenario.
\end{abstract}
\maketitle

\section{Introduction}

Recent years have seen an enormous progress in preparing, controlling and probing 
ultra cold atomic gases loaded in optical lattices \cite{Bloch_rmp08}.
Their high degree of tunability allows to change in time the microscopic parameters controlling
interactions among atoms and to measure the resulting quantum evolution. 
At the same time their excellent isolation from the environment makes those systems 
particularly well suited to address questions related to non equilibrium phenomena in isolated 
many body quantum systems. These major achievements triggered a huge interest on time dependent 
phenomena in condensed matter systems. In this respect, the recent experimental realization 
of a fermionic Mott insulator~\cite{Fermionic_Mott_08,Fermionic_Mott_08_Bloch} opened the way to investigate 
out-of-equilibrium phenomena in strongly correlated fermionic systems \cite{Demler_doublons_prl10}.

From a more theoretical perspective these experiments offer the chance to probe 
strongly correlated systems in a completely novel regime. Indeed, when driven out of equilibrium,
interacting quantum systems can display peculiar dynamical behaviors or even be trapped into metastable 
configurations that differ completely from their equilibrium counterpart~\cite{Rosch_prl08,Rosch_prl10}. 
Although actual experiments are always performed by tuning parameters at a finite rate, 
an useful idealization consists in a so called quantum quench \cite{Calabrese_Cardy}. 
Here the system is firstly prepared in the many-body ground state $\vert\Psi_i\rangle$ 
of some initial Hamiltonian $H_i$ which is then \emph{suddenly} changed to $H_f\neq H_i$, for 
example by globally switching on or off some coupling constants.
As a consequence of this instantaneous change the initial state $\vert\Psi_i\rangle$ turns 
to be an highly excited state of the final Hamiltonian. 
Naturally, many non trivial questions arise concerning the 
real-time evolution after the quantum quench. The interest on these classes of non equilibrium problems relies 
both on the dynamics itself, \cite{Corinna_prl07,Peter_prl09} as well as on the long-time properties where the question 
of thermalization or its lack of is still 
highly debated.\cite{Olshanii_Rigol_nature,Biroli_Corinna_prl10,Gogolin_prl11}
This issue is not only of fundamental theoretical interest 
but also of great practical relevance for establishing 
whether and to what extent experiments on cold atoms 
could reproduce equilibrium phase diagrams of model hamiltonians. 

The literature on quantum quenches in interacting bosonic and fermionic systems is by now very broad, 
see for example the recent topical reviews \cite{Review_FlowEq_DMFT,Review_MPS,CazalillaRigol,SilvaRMP}
For what concerns strongly correlated  electrons in more than one dimension,
the subject is still largely unexplored and progresses have been done only very recently. 
The single band Hubbard model \cite{Hubbard,Gutzwiller_PRL,Kanamori} represent one of the simplest yet 
non trivial models encoding the physics of strong correlations, namely the competition between electronic
wave function delocalization due to hopping $t$ and charge localization due to large Coulomb repulsion $U$.
Its Hamiltonian reads
\be\label{eqn:hubb}
\m{H}\left(t\right)= -\sum_{\sigma}\sum_{\langle \bR,\bRp\rangle}\,t_{\bR\bRp}\,
c^{\dagger}_{\bR\sigma}\,c^\dagga_{\bRp\sigma}+U\left(t\right)\sum_{\bR}\,
n_{\bR\uparrow}\,n_{\bR\downarrow}\,.
\ee
Despite the everlasting interest on its groundstate properties, theoretical investigations on the non 
equilibrium dynamics of this paradigmatic strongly correlated model have been started only very recently.
The dynamics of Fermi system after a sudden switch-on of the Hubbard interaction has been studied firstly in 
Refs.~\onlinecite{Kehrein_prl08,Moeckel2009} using the flow-equation approach and then in Refs.~\onlinecite{Werner_prl09,Werner_long10}
using Non equilibrium Dynamical Mean Field Theory (DMFT). 
Results suggest the existence of two different regimes 
in the real-time dynamics, depending on the
final interaction strength $U_f$. At weak coupling,\cite{Kehrein_prl08} the systems is trapped at long-times into a 
quasi-stationary regime which looks as a zero temperature Fermi Liquid from the energetic point of view but 
features a non thermal distribution function in which correlations are more effective than in equilibrium.
This \emph{pre-thermalization} phenomenon has been confirmed by DMFT results,\cite{Werner_prl09} which further indicate 
a true dynamical transition above a critical $U_{fc}$ towards another regime with pronounced oscillations 
in the dynamics of physical quantities. This picture has been recently confirmed by means of a simple and flexible 
approximation scheme based on a proper extension of the Gutzwiller variational method~ \cite{SchiroFabrizio_prl10}.
Results for the time depedent mean field theory show at half filling, a sharp transition between two 
different regimes of small and large coherent oscillations, separated by a critical line of quenches where the system finds
a fast way to relax. Away from particle hole symmetry the transition is washed out, leaving a sharp 
crossover visible in the dynamics and in the long-time averages of physical quantities.

The aim of the present work is twofold. From one side, 
we present details on the time dependent Gutzwiller method 
for fermions and discuss its application to the problem of an interaction quench in the single band Hubbard model. 
Secondly, we discuss the role of quantum fluctuations on top of the Gutzwiller dynamics. In order to do that
we formulate the original Hubbard model in terms of an auxilary 
Quantum Ising Model in a transverse field coupled to 
free fermionic quasiparticle. Such a $Z_2$ slave spin theory, 
introduced in Refs.~\onlinecite{Z2-1,Z2-2} for the equilibrium problem, 
allows us to study the effect of small quantum fluctuations, 
both in equilibrium as well as for the non equilibrium dynamics.
We notice that the role of quantum fluctuations on this mean field \emph{dynamical} transition is of broader theoretical
interest, as recent investigations have shown the very same phenomenon occurs in other models of interacting quantum field
theories~\cite{SciollaBiroli_prl10,GambassiCalabrese_preprint}.

The paper is organized as follows. In the first part we introduce the time dependent variational method we have devised
to describe non equilibrium dynamics in correlated electrons systems. Section \ref{sect:general} 
is devoted to a general formulation while section \ref{sect:hubbard} to the study of quantum quenches in the single band
fermionic Hubbard Model. In the second part of the paper we broaden the perspective and formulate the 
Hubbard model in terms of auxiliary Quantum Ising Model coupled to free fermionic quasiparticles.
In section \ref{sect:Ising} we show how the mapping works and how to recover the Gutzwiller results. 
Section \ref{sect:fluct} is devoted to the role of quantum fluctuations. 
Finally section \ref{sect:conclusion} is for conclusions.

\section{A General Formulation}\label{sect:general}

We assume a system of interacting electrons that is initially in a state with many-body 
wavefunction $|\Psi_0\rangle$. 
For times $t>0$, $|\Psi_0\rangle$ is let evolve with the Hamiltonian $\mathcal{H}$, which could even be 
explicitly time-dependent. We shall assume that short range correlations are strong either in 
the initial wavefunction, or in $\mathcal{H}$, or in both. 
The goal is calculating average values of operators during the time evolution.   
Because of interaction, a rigorous calculation is unfeasible, so that an approximation scheme 
is practically mandatory. Our choice will be to use a proper extension of the Gutzwiller wavefunction and 
approximation, which is known to be quite effective at equilibrium when strong short-range 
correlations are involved. 

We start by defining a class of many-body wavefunctions of the form  
\bea
|\Psi(t)\rangle &=& 
\prod_\mathbf{R} \mathrm{e}^{-i\mathcal{S}_\bR(t)}\, \m{P}_\bR(t) \,|\Phi(t)\rangle \nonumber \\
&\equiv& \mathcal{P}(t)\,|\Phi(t)\rangle,\label{eqn:ansatz}
\eea
where $|\Phi(t)\rangle$ are time-dependent variational wavefunctions for which Wick's theorem holds, 
hence Slater determinants or BCS wavefunctions, while   
$\m{P}_\bR(t)$ and $\m{S}_{\bR\alpha}$ are hermitian operators that act on the Hilbert space at site $i$ 
and depend on the variables $\lambda_{\bR\alpha}(t)$ and $\phi_{\bR\alpha}(t)$:    
\bea
\m{P}_\bR(t) &=& \sum_{\bR\alpha}\,\lambda_{\bR\alpha}(t)\,\m{O}_{\bR\alpha},\label{eqn:def_amplitude}\\
\fract{\partial}{\partial \phi_{\bR\alpha}}\,\mathrm{e}^{-i\m{S}_\bR} &=& 
-i\,\m{O}_{\bR\alpha}\,\mathrm{e}^{-i\m{S}_\bR},\label{eqn:def_phase}
\eea
where $\m{O}_{\bR\alpha}$ can be any local hermitian operator. It follows that the average value 
of $\m{O}_{\bR\alpha}$
\be
O_{\bR\alpha} = \langle \Psi(t)|\,\m{O}_{\bR\alpha}\,|\Psi(t)\rangle,\label{def:O_ialpha}
\ee
is a functional of all the variational parameters. We shall assume that it is possible to invert \eqn{av-value-local} 
and express the parameters $\lambda_{\bR\alpha}$ as functionals of all the $O_{\bRp\beta}$, $\phi_{\bRp\beta}$ 
as well as of the parameters that define $|\Phi(t)\rangle$.

Since $|\Psi(t)\rangle$ spans a sub-class of all possible many-body wavefunctions, in 
general it does not solve the Schr{\oe}dinger equation but can be chosen to be as close as possible 
to a true solution. This amounts to search for the saddle point of the functional 
\be
\m{S}[\Psi^{\dagger},\Psi] = \int\,dt\,\langle\Psi(t)\vert\,i\partial_t-\m{H}\vert\Psi(t)\rangle,
\label{def:S-action}
\ee
with $|\Psi(t)\rangle$ of the form as in Eq.~\eqn{eqn:ansatz}. The Gutzwiller approximation gives a 
prescription for calculating $\m{S}$, which is exact in infinite coordination 
lattices,\cite{Gebhard,Michele_PRB07_dimer} although it is believed to provide reasonable results also when the 
coordination is finite. We impose that 
\bea
\langle \Phi(t)|\,\m{P}_\bR^2(t)\,|\Phi(t)\rangle &=& 1,\label{Gutzwiller-cond-1}\\
\langle \Phi(t)|\,\m{P}_\bR^2(t)\,\m{C}_{\bR\alpha}\,|\Phi(t)\rangle &=& 
\langle \Phi(t)|\,\m{C}_{\bR\alpha}\,|\Phi(t)\rangle,\label{Gutzwiller-cond-2}
\eea
where $\m{C}_{\bR\alpha}$ is any bilinear form of the single-fermion operators at site $\bR$, 
$c^\dagger_{\bR a}$ and $c^\dagga_{\bR a}$ with $a$ the spin/orbital index.

Within the Gutzwiller approximation and provided Eqs.~\eqn{Gutzwiller-cond-1} 
and \eqn{Gutzwiller-cond-2} hold, the average value of any local operator $\m{O}_{\bR\alpha}$ is assumed to 
be\cite{Michele_PRB07_dimer}
\bea
O_{\bR\alpha} &=& \langle \Psi(t)|\,\m{O}_{\bR\alpha}\,|\Psi(t)\rangle \label{av-value-local}=\\
&=& 
\langle \Phi(t)|\,\m{P}_\bR(t)\,\mathrm{e}^{i\m{S}_\bR(t)}\,\m{O}_{\bR\alpha}\,\mathrm{e}^{-i\m{S}_\bR(t)}\,
\m{P}_\bR(t)\,|\Phi(t)\rangle,\nonumber 
\eea
which can be easily computed by the Wick's theorem. Seemingly, given two local operators, 
$\m{O}_{\bR\alpha}$ and $\m{O}_{\bRp\beta}$ at different sites $\bR\not=\bRp$, the following expression is 
assumed
\begin{widetext}
\be
\langle \Psi(t)|\,\m{O}_{\bR\alpha}\,\m{O}_{\bRp\beta}\,|\Psi(t)\rangle =
\langle \Phi(t)|\,\m{P}_\bR(t)\,\mathrm{e}^{i\m{S}_\bR(t)}\,\m{O}_{\bR\alpha}\,\mathrm{e}^{-i\m{S}_\bR(t)}\,
\m{P}_\bR(t)\,
\m{P}_\bRp(t)\,\mathrm{e}^{i\m{S}_\bRp(t)}\,\m{O}_{\bRp\beta}\,\mathrm{e}^{-i\m{S}_\bRp(t)}\,
\m{P}_\bRp(t)\,|\Phi(t)\rangle,\label{av-value-non-local}
\ee
\end{widetext}
which can be also readily evaluated. For consistency, one should keep 
only the leading terms in the limit of infinite coordination lattices.\cite{Michele_PRB07_dimer} 
For instance, if $|\Phi(t)\rangle$ is a Slater determinant and $\m{O}_{\bR\,a} = c^\dagger_{\bR\,a}$ 
while $\m{O}_{\bRp\,b}=c^\dagga_{\bRp\,b}$, then   
\bea
&&\langle \Psi(t)|\,c^\dagger_{\bR\,a}\,c^\dagga_{\bRp\,b}\,|\Psi(t)\rangle =\nonumber \\
&&=\sum_{c d}\, Q^*_{\bR,ac}\,Q^\dagga_{\bRp,bd}\, 
\langle \Phi(t)|\,c^\dagger_{\bR\,c}\,c^\dagga_{\bRp\,d}\,|\Phi(t)\rangle, \label{def:cdaggeri-cdaggaj}
\eea
where the matrix elements $Q_{\bR,ab}$ are obtained by solving
\bea
&& \langle \Phi(t)|\,\m{P}_\bR(t)\,\mathrm{e}^{i\m{S}_\bR(t)}\,c^\dagger_{\bR\,a}\,
\mathrm{e}^{-i\m{S}_\bR(t)}\,
\m{P}_\bR(t)\,c^\dagga_{\bR\,c}\,|\Phi(t)\rangle \nonumber \\
&&~~= 
\sum_b\,Q^*_{\bR,ab}\,\langle \Phi(t)|\,c^\dagger_{\bR\,b}\,c^\dagga_{\bR\,c}\,|\Phi(t)\rangle.
\label{def-R}
\eea  

Within the Gutzwiller approximation one finds that 
\be
i\langle \Psi(t)|\partial_t\Psi(t)\rangle = \sum_{\bR\alpha}\,\dot{\phi}_{\bR\alpha}\,O_{\bR\alpha} 
+i\langle \Phi(t)|\partial_t\Phi(t)\rangle,\label{diff-t}
\ee
so that 
\bea
\label{eqn:action}
\m{S}[\Psi^{\dagger},\Psi] 
&=& \int\, dt\, \bigg( \sum_{\bR\alpha}\, \dot{\phi}_{\bR\alpha}\,O_{\bR\alpha} 
- E\left[\phi_{\bR\alpha},O_{\bR\alpha},\Phi\right] \nonumber\\
&& \phantom{ \int\, dt\, \sum_{\bR\alpha}\,} +i \langle \Phi(t)|\partial_t\,\Phi(t)\rangle\bigg),
\eea  
where 
\bea 
E\left[\phi_{\bR\alpha},O_{\bR\alpha},\Phi\right] &=& \langle \Phi(t)|\,\m{H}_*\,|\Phi(t)\rangle,\label{def:H-cassical}
\\
\m{H}_* &=& P^{\dagger}(t)\,\m{H}\,P(t)\label{def:H*}\,.
\eea
The saddle point of $\m{S}$ in Eq.~\eqn{eqn:action} with respect to $\phi_{\bR\alpha}$ 
and $O_{\bR\alpha}$ is readily obtained by imposing
\bea
\dot{\phi}_{\bR\alpha} &=& \fract{\partial E}{\partial O_{\bR\alpha}},\label{eqn:eom_1}\\ 
\dot{O}_{\bR\alpha} &=& -\fract{\partial E}{\partial \phi_{\bR\alpha}},\label{eqn:eom_2}
\eea
showing that these pairs of variables act like classical conjugate fields with Hamiltonian $E$. 
As far as $|\Phi(t)\rangle$ is concerned, since it is either a Slater determinant or a BCS wavefunction, 
the variation with respect to it leads to similar equations as 
in the time-dependent Hartree-Fock approximation,\cite{NegeleOrland_book} namely, in general, non-linear single particle 
Schr{\oe}dinger equations. 

In conclusion, the variational principle applied to the Schr{\oe}dinger equation and combined with the 
Gutzwiller approximation amounts to solve a set of equations that is only slightly more complicated 
than the conventional time-dependent Hartree-Fock approximation, yet incomparably simpler 
than solving the original Schr{\oe}dinger equation. We note that, in the above scheme, the Gutzwiller 
variational parameters $\lambda_{\bR\alpha}$ in Eq.~\eqn{eqn:def_amplitude}, or better $O_{\bR\alpha}$ 
in Eq.~\eqn{def:O_ialpha}, 
have their own dynamics because of the presence of their conjugate fields $\phi_{\bR\alpha}$. This marks the 
difference with the time-dependent variational scheme introduced by Seibold and Lorenzana\cite{Lorenzana}, 
where the time evolution of $\lambda_{\bR\alpha}$ is only driven by the time evolution of the 
Slater determinant. We shall see that this difference may play an important role.

\section{Quantum Quenches in the Hubbard Model}\label{sect:hubbard}

We now turn to the problem of our interest and discuss the non equilibrium dynamics 
in the Hubbard model (\ref{eqn:hubb}) using the time dependent variational scheme introduced above. 
This calculation allows to benchmark the method towards more reliable techniques, a compulsory step before 
moving to more complicated situations where rigorous results are lacking.   
In particular we shall study the dynamics after a sudden change of the local interaction, 
starting from the zero-temperature {\sl variational} 
ground state with $U(t\leq 0)=U_i$ then quenching the interaction to $U(t>0)=U_f$.
Notice that since the initial state is described within the equilibrium Gutzwiller approximation, 
which provides a poor description of the Mott Insulator, we have to restrict our analysis to 
strongly correlated yet metallic initial conditions, namely to $U_i<U_c$ where $U_c$ is the critical 
interaction strength for the Mott transition within the Gutzwiller approximation. 
Moreover, in what follows we shall completely disregard magnetism, 
considering only paramagnetic and homogeneous wave functions.

\subsection{Time Dependent Gutzwiller Approximation}

We take $\mathcal{H}$ to be the single band Hubbard model (\ref{eqn:hubb}) 
and assume a correlated time-dependent wave function of the form (\ref{eqn:ansatz}) with
\bea \label{eqn:P_i}
\m{P}_\bR(t) &=& \sum_{n=0}^2\,\lambda_{\bR,n}(t)\,\m{P}_{\bR,n}\,,\\
\label{eqn:S_i}
\m{S}_\bR(t) &=& \sum_{n=0}^2\,\phi_{\bR,n}(t)\,\m{P}_{\bR,n}\,,
\eea
where $\m{P}_{\bR,n}$ is the projector at site $\bR$ onto configurations with $n=0,\dots,2$ electrons.
Notice that equations (\ref{eqn:P_i}-\ref{eqn:S_i}) imply that $\phi_{\bR,n}(t)$ plays the role 
of the conjugate variable of 
\be
P_{\bR,n} = \langle \Psi(t)\vert \m{P}_{\bR,n}\vert \Psi(t)\rangle.\label{def:occ-prob} 
\ee
For non-magnetic wavefunctions, the renormalization parameters in Eq.~(\ref{def-R}) do 
not depend on the spin index and read
\bea \label{eqn:Z_qp_general}
Q_i &=&\frac{\sqrt{P_{\bR,1}}}{\sqrt{n_\bR\left(1-n_\bR/2\right)}}\,
\Bigg(\sqrt{P_{\bR,2}}\;e^{i\left(\phi_{\bR,2}-\phi_{\bR,1}\right)}\nonumber \\
&& \phantom{\frac{\sqrt{P_{\bR,1}}}{\sqrt{n_\bR(1-n_\bR/2)}}
} +
\sqrt{P_{\bR,0}}\;e^{i\left(\phi_{\bR,1}-\phi_{\bR,0}\right)}\Bigg),
\eea
where 
\[
n_\bR = \sum_\sigma\,\langle \Phi(t)|c^\dagger_{\bR\sigma}c^\dagga_{\bR\sigma}|\Phi(t)\rangle,
\]
is the average on-site occupancy. The two constraints Eqs.~\eqn{Gutzwiller-cond-1} and 
\eqn{Gutzwiller-cond-2} imply that the quantities $P_{\bR,n}$ in \eqn{def:occ-prob} behave 
as genuine occupation probabilities with 
\ba
\sum_n\,P_{\bR,n} &=& 1,\\
\sum_n\,n\,P_{\bR,n} &=& n_\bR.
\ea
If we set $P_{\bR,2}\equiv D_\bR$ then $P_{\bR,0}=1-n_\bR+D_\bR$ and $P_{\bR,1}=n_\bR-2D_\bR$. 
We also assume that $\phi_{\bR,0}=\phi_{\bR,2}=\phi_\bR$ while $\phi_{\bR,1}=0$, so that the 
energy functional $E$ becomes  
\bea 
\label{eqn:gutzwiller_energy}
E\left[\phi_\bR,D_\bR,\Phi\right] &=& \langle\Psi(t)\vert\m{H}\vert\Psi(t)\rangle=U_f\sum_{\bR}\,D_{\bR} +\nonumber\\
&& 
+\sum_{\langle \bR\bRp\rangle}\,Q_\bR^\dagga\,Q_\bRp^*\, w_{\bR\,\bRp}(t) + H.c.\, ,
\eea
where 
\be
w_{\bR\bRp}(t)=t_{\bR\,\bRp}\,\sum_\sigma\,\langle 
\Phi(t)\vert\,c^{\dagger}_{\bR\sigma}c^\dagga_{\bRp\sigma}\,\vert\Phi(t) \rangle,\label{def:w}
\ee
while $Q_{\bR}(t)$ defined in equation (\ref{eqn:Z_qp_general}) reads
\bea\label{eqn:Z_qp_i}
Q_\bR = \sqrt{\frac{n_\bR-2D_\bR}{n_\bR\left(1-n_\bR/2\right)}}\times\phantom{\quad\quad\quad}\nonumber\\\;
\phantom{\quad\quad\quad}\times\left(\sqrt{D_\bR+1-n_\bR}\;e^{i\phi_\bR}+\sqrt{D_\bR}\;e^{-i\phi_\bR}\right)\,.
\eea
By the variational energy (\ref{eqn:gutzwiller_energy}) we can readily obtain the equations of motion for
the double occupancy $D_\bR$ and its conjugate variable $\phi_\bR$ using (\ref{eqn:eom_1},\ref{eqn:eom_2}). 
In addition, the dynamics of these variational parameters is further coupled to a time dependent 
Schroedinger equation for the Slater determinant. If this latter is initially homogeneous, 
then translational symmetry is mantained during the time evolution, hence $Q_\bR(t)=Q(t)$ independent of $\bR$. 
Moreover, if the Slater determinant $|\Phi(t=0)\rangle$ 
is initially the Fermi sea, i.e. the lowest energy eigenstate of the hopping Haimiltonian, then 
its time evolution caused by the time dependent hopping $\vert\,Q(t)\vert^2\,t_{\bR\bRp}$ becomes trivial
\[
|\Phi(t)\rangle = \exp\bigg(-i\,V\,\bar{\epsilon}_n\,\int_0^t d\tau\,\vert\,Q(\tau)\vert^2\bigg)\;|\Phi(t)\rangle,
\]
where $\bar{\epsilon}_n$ is the average energy per site of the hopping Hamiltonian with electron density 
$n$ on a lattice with $V$ sites. In other words, the matrix elements $w_{\bR\bRp}(t)$ in Eq.~\eqn{def:w} 
are in this case time independent.

\subsection{Saddle-point equations}

In conclusion, within the Gutzwiller approximation and assuming a 
homogeneous and non-magnetic wavefunction 
the classical Hamiltonian (\ref{eqn:gutzwiller_energy}) for the single degree of freedom $D_\bR\equiv\,D$ and its 
conjugate variable $\phi_\bR\equiv\,\phi$ reads
\be\label{eqn:H_class_onesite}
E[D,\phi]= U_f D(t)+\bar{\eps}_n\,Z(D,\phi)\,,
\ee
where we remind that $\bar{\eps}_n$ is the average hopping energy of a Fermi sea 
with density $n=1-\delta$ while $Z=\vert\,Q\vert^2$ is the effective 
quasiparticle weight, which reads from equation (\ref{eqn:Z_qp_i})
\bea\label{eqn:Z_qp_onesite}
Z\left(D,\phi\right) &=& \frac{2\left(n-2D\right)}{n\left(2-n\right)}\times\\
&&\left[\left(\sqrt{D+\delta}-\sqrt{D}\right)^2+4\cos^2\phi\sqrt{D}\sqrt{D+\delta}\right]\,.\nonumber
\eea
Notice that $Z$ does not depends only from the double occupation $D$, as one would expect in equilibrium, but features 
a dependence from the phase $\phi$ which is crucial in order to induce a non trivial dynamics.

The classical equations of motion for this integrable system immediately follow 
from (\ref{eqn:H_class_onesite}) 
\bea\label{eqn:pendulum1}
\dot{\phi} &=& \frac{U_f}{2} +\frac{\bar{\eps}_n}{2}\,\frac{\partial Z}{\partial D}\,,\\
\label{eqn:pendulum2}
\dot{D} &=& -\frac{\bar{\eps}_n}{2}\, \frac{\partial Z}{\partial\phi}\,,
\eea
In the following we will use the MIT critical interaction at half-filling, 
$U_c=-8\bar{\eps}_{n=1}\equiv -8\bar{\epsilon}$, as the basic unit of energy and 
define accordingly the dimensionless quantities $u_f=U_f/U_c$ and $u_i=U_i/U_c$, as well 
a dimensionless time $t = t\,U_c$.  
In addition we shall assume for simplicity a flat density of states so that $\bar{\epsilon}_n = 
n(2-n)\,\bar{\epsilon} = -n(2-n)/8\, U_c$.   

The initial conditions for the classical dynamics (\ref{eqn:pendulum1})-(\ref{eqn:pendulum2}) read
\be
D(t=0)=D_i\,,\qquad\,\phi(t=0)=0\,,
\ee
where $D_i$ is the equilibrium zero temperature double occupancy for interaction 
$u_i$ and doping $\delta$ that can be
easily computed from an equilibrium Gutzwiller calculation, which is nothing but annihilating the 
right hand sides of Eqs.~\eqn{eqn:pendulum1} and \eqn{eqn:pendulum2} with interaction $U_i$ instead of $U_f$. 

\begin{figure}[t]
\begin{center}
\epsfig{figure=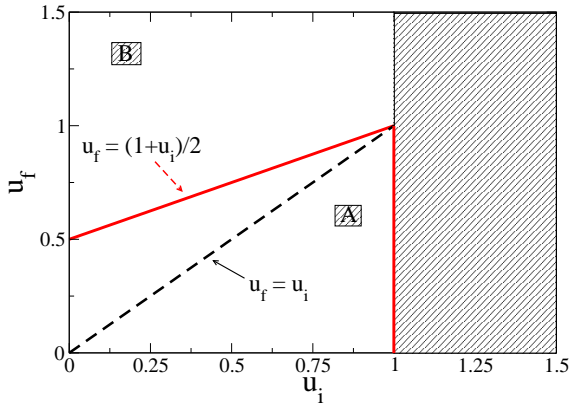,scale=0.3}
\caption{Sketch of the \emph{phase diagram} in the $u_i,u_f$ plane for the quench dynamics of the single band Hubbard model
within the Gutzwiller approximation at half-filling. Two different dynamical regimes corresponding to weak and strong 
coupling dynamics (A and B in the plot) are found depending whether the final interaction $u_f$ lies above or below the
 critical quench line $u_{fc}=\frac{1+u_i}{2}$. For quantum quenches along this line the dynamics features an exponential
 relaxation toward a steady state.}
\label{fig:fig1}
\end{center}
\end{figure}

It is worth noticing that, apart from the trivial case in which $u_f=u_i$, the classical dynamics 
(\ref{eqn:pendulum1})-(\ref{eqn:pendulum2}) admits a non-trivial stationary solution $D=0$ and $\cos^2\phi=u_f$,
 which is compatible with the initial conditions only at half-filling and  $u_f=u_{fc}=(1+u_i)/2$. 
It turns out that $u_{fc}$ identifies a dynamical critical point that separates two different
regimes similarly to a simple pendulum. When $u_f<u_{fc}$,  $2\phi(t)$ oscillates around
the origin, while, for $u_f>u_{fc}$, it performs a cyclic motion around the whole circle.
In order to characterize the different regimes, we focus on three physical quantities,
the double occupancy $D(t)$, the quasiparticle residue $Z(t)$ and their period of
oscillation, $\m{T}$. 

Before discussing in some detail the results of the classical dynamics 
(\ref{eqn:pendulum1})-(\ref{eqn:pendulum2}), 
it is useful to cast it into a closed first-order differential equation for one of the two conjugate 
variables $D$ or $\phi$. Indeed the dynamics conserves the energy, namely
\be\label{eqn:integral_of_motion}
E(t)=u_f D(t) - \fract{n(2-n)}{8}\,Z(t)\equiv E_0
\,,\qquad t>0
\ee
where $E_0$ is the total energy soon after the quench, which reads
\be\label{eqn:initial_energy}
E_0 = u_f\,D_i
- \fract{n(2-n)}{8}\,Z_{i}\,.
\ee
with $Z_i$ the equilibrium zero temperature quasiparticle weight for interaction 
$u_i$ and doping $\delta$. The simplest way to proceed is to eliminate $\phi$ from Eq.(\ref{eqn:integral_of_motion}) 
in favor of the double occupancy $D(t)$.
From Eq.(\ref{eqn:Z_qp_onesite}) we obtain
\be\label{eqn:effective1}
\cos^2\phi = - \frac{E_0-u_fD+\left(n-2D\right)\left(\sqrt{D+\delta}-\sqrt{D}\right)^2/4}
{\left(n-2D\right)\sqrt{D\left(D+\delta\right)}},
\ee
which can be inserted into (\ref{eqn:pendulum2}) and leads, after some algebra, to the equation of motion
\be\label{eqn:gen_dyn_docc}
\dot{D} = \pm\sqrt{\Gamma(D)}\,.
\ee
Here $\Gamma(D)$ can be thought as an effective potential controlling the dynamical behavior of $D(t)$. 
We note that, since the problem is one dimensional,  many properties of the solution (\ref{eqn:gen_dyn_docc}) 
can be inferred directly from the knowledge of $\Gamma(D)$, without explicitly solving the dynamics.
In the next two sections we will discuss in detail the structure of this solution, considering both
the half filled and the doped case.

%

\subsection{Quench Dynamics at Half-Filling}\label{sect:half_filling}

We start by considering half-filling, i.e. $\delta=0$, and for simplicity we
fix $u_f>u_i$, see figure \ref{fig:fig1}. As we already anticipated, 
the dynamical behavior of the system changes drastically when the final value of the interaction $u_f$
crosses the critical line $u_{fc}\equiv\left(1+u_i\right)/2$. 

The existence of such a line of critical values clearly emerges from the structure of the effective potential 
$\Gamma\left(D\right)$ and in particular from the behavior of its positive roots, 
which are the inversion points of the one dimensional motion~(\ref{eqn:gen_dyn_docc}). 

\begin{figure}[t]
\begin{center}
\epsfig{figure=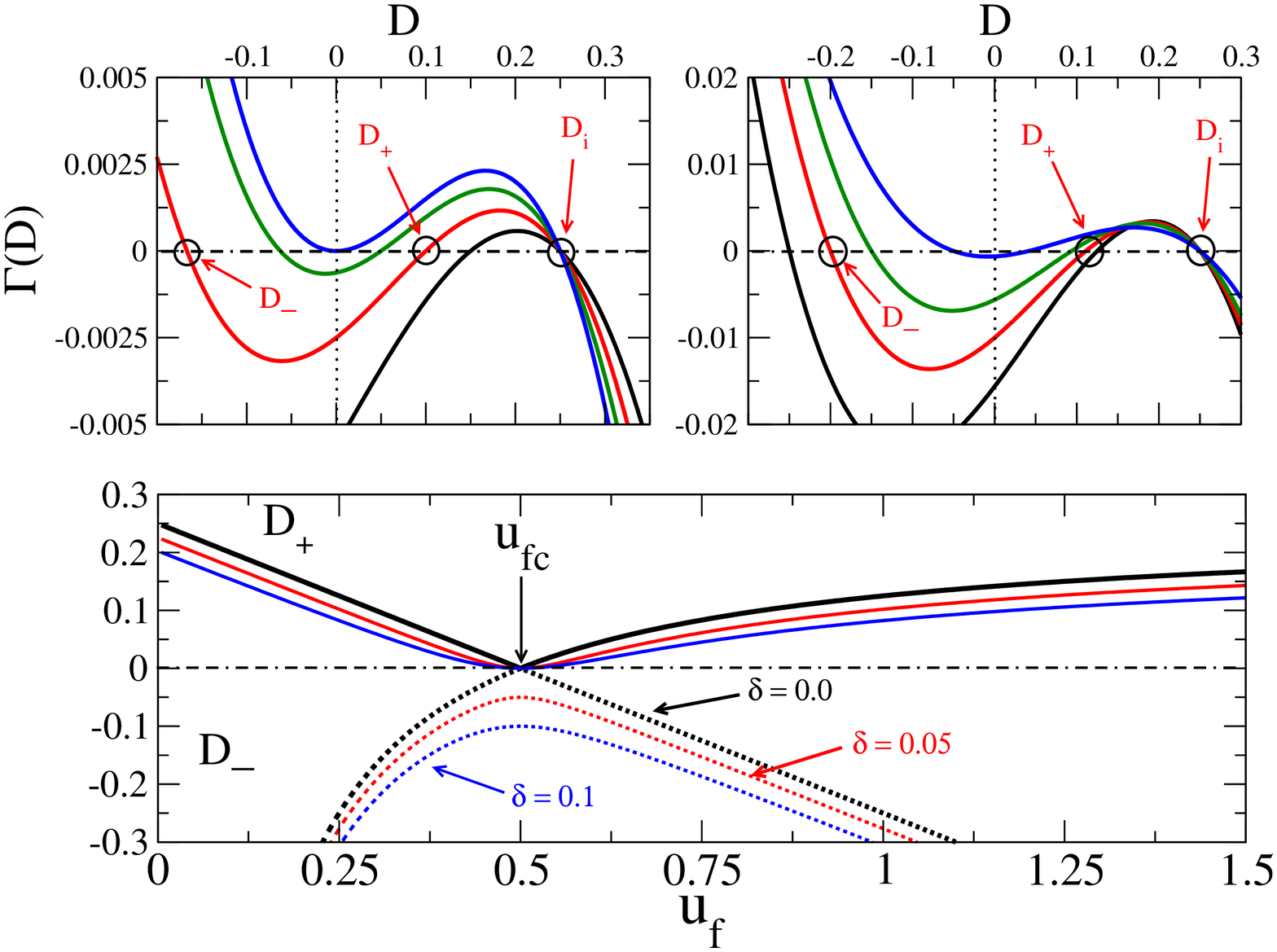,scale=0.325}
\caption{Top Panel: Effective potential $\Gamma\left(D\right)$ for $u_f=0.2,0.3,0.4,0.5$ (right) and $u_f=0.6,0.8,0.9,1.0$
(left). Bottom panel: Inversion points $D_{+},D_{-}$ as a function of $u_f$ at fixed $u_i=0$ for zero and finite doping.}
\label{fig:fig2}
\end{center}
\end{figure}

As one can see from figure \ref{fig:fig2}, $\Gamma(D)$ has three simple zeros, two of them being positive.
It turns out that the equilibrium Gutzwiller solution $D_i$ is always one of the roots of
the effective potential, for any $u_f$, see figure \ref{fig:fig2} (top panels).
The remaining two, $D_{\pm}$, depend strongly on $u_f$ as we show in the bottom panel of figure \ref{fig:fig2}. 
Since the one dimensional motion is constrained to the interval $[D_+,D_i]$, where $\Gamma(D)$ is positive,
we expect to find periodic solution of the dynamics (\ref{eqn:gen_dyn_docc}). However, the properties of this solution
will largely depend on the behavior of $D_+$ as a function of $u_f$. 
As we see, $D_+$ first decreases linearly with $u_f$, vanishing at $u_{fc}$ where it becomes degenerate with $D_-$ 
(see figure \ref{fig:fig2}). Then for $u_f>u_{fc}$ it starts increasing again, approaching $D_i$ 
in the infinite quench limit. It turns out that $D_+$ has a simple form, which reads
\be\label{eqn:Dplus}
D_{+}=\left\{
\begin{array}{ll}
u_f < u_{fc}\,\, & \left(u_{fc}-u_f\right)/2\\
u_f > u_{fc}\,\, & D_i\left(1-\frac{u_{fc}}{u_f}\right)\\
\end{array}
\right.
\ee
Two different dynamical behaviors are therefore expected as a result of this peculiar dependence. In addition, due to
the degeneracy of simple roots occurring at $u_f=u_{fc}$ we expect here a special trajectory, where relaxation to a steady
state can exist. This qualitative picture is confirmed by the actual solution of the classical dynamics (\ref{eqn:gen_dyn_docc}),
whose results we are going to present, both for weak ($u_f<u_{fc}$) and strong ($u_f>u_{fc}$) quantum quenches.

\subsubsection*{Weak Quenches: $u_f<u_{fc}$}

\begin{figure}[t]
\begin{center}
\epsfig{figure=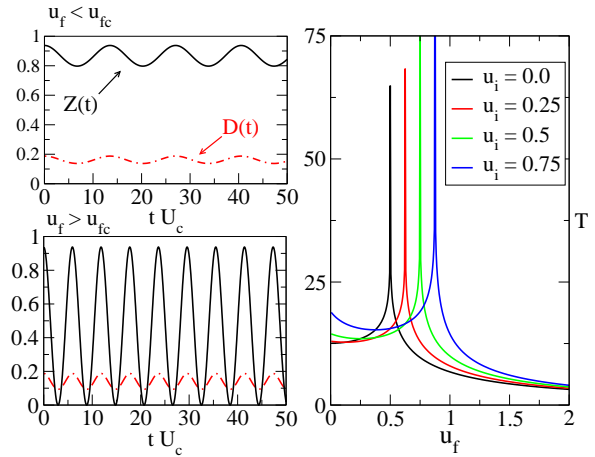,scale=0.325}
\caption{Left Panel: Mean-field dynamics for quantum quenches to $u_f$ below (top) and above (bottom) the critical line.
 Right Panel: Period of oscillation $\m{T}_D$ as a function of $u_f$ for $u_i=0.0,0.25,0.5,0.75$. 
Notice the log-singularity at $u_{fc}$.}
\label{fig:fig3}
\end{center}
\end{figure}

For weak quantum quenches to $u_f<u_{fc}$, the dynamics of both double occupation $D(t)$ and quasiparticle weight $Z(t)$
shows coherent \emph{oscillations}, see figure \ref{fig:fig3}, which do not die out.
The lack of relaxation toward a steady state is clearly an artifact of 
our semiclassical approach that does not account for  
quantum fluctuations. This is particularly true for weak quenches 
starting from the gapless metallic phase,
where fast damping of the oscillations is expected due to the available 
continuum of low-lying excitations.

\begin{figure}[t]
  \begin{center}
  \epsfig{figure=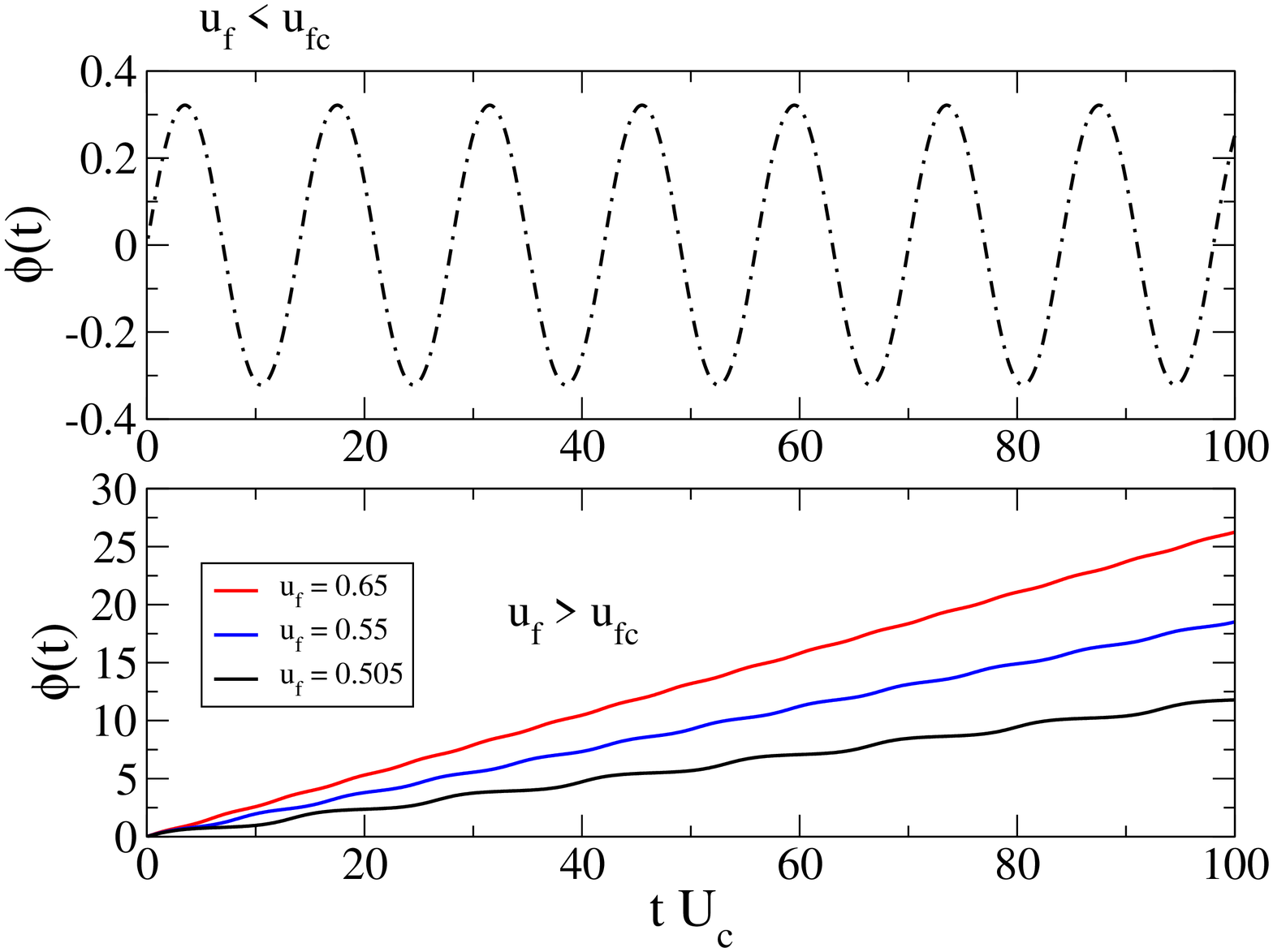,scale=0.325}
  \caption{Dynamics of the phase in for quenches below and above the critical value $u_{fc}$. 
Notice that for small quenches the phase oscillates around zero while for $u_f>u_{fc}$ the dynamics is no more bounded 
since the energy is sufficient to overcome the potential barrier.}
  \label{fig:fig4}
  \end{center}
  \end{figure}

Although oversimplified, the dynamics in the weak quench limit contains some interesting features that are worth to discuss.
In particular, we focus on the period $\m{T}$ of the coherent oscillations as a function of the final
interaction $u_f$. It is easy to see that $\m{T}$ is given by 
\be\label{eqn:period_wc}
\m{T}= 2\,\int_{D_{+}}^{D_{i}}\,\frac{dD}{\sqrt{\Gamma\left(D\right)}}
=\frac{4\sqrt{2}\,K\left(k\right)}{\sqrt{Z_i}}\,,
\ee
where $K\left(k\right)$ is the complete elliptic integral of the first kind with argument 
$k^2=4u_f\left(u_f-u_i\right)/Z_i$.
As we show in the right panel of figure \ref{fig:fig3}, upon increasing $u_f$ 
the period $\m{T}$ grows eventually diverging logarithmically as the critical quench line $u_f=u_{fc}$ is approached.
This can be seen explicitly in Eq.(\ref{eqn:period_wc}). 
Indeed, for $u_f\rightarrow u_{fc}$, the argument of the complete elliptic integral approaches $k=1$
\be
1-k^2 = \left(u_{fc}-u_f\right)
\left(1+\frac{u_f}{2\,u_{fc}\,D_i}\right)\,.
\ee
Therefore, using the known asymptotic result
 $K\left(k\right)\simeq \log\left(4/\sqrt{1-k^2}\right)$ we find 
\be
\m{T}\sim \frac{4}{\sqrt{1-u_i^2}}\,\log\left(\frac{1}{u_{fc}-u_f}\right)\,.
\ee
Such a diverging time scale signals a sharp transition to a completely different dynamical regime for $u_f>u_{fc}$. 
Before moving to this strong coupling regime we briefly discuss the dynamics of the phase $\phi(t)$ in the weak quench case,
which can be easily obtained by eliminating the double occupation $D(t)$ from the original system 
(\ref{eqn:pendulum1}-\ref{eqn:pendulum2}). 
As shown in figure \ref{fig:fig4}, in the present weak 
quench regime ($u_f<u_{fc}$) the phase oscillates around 
the \emph{equilibrium} fixed point $\phi=0$, 
with the same period $\m{T}$. As we are going to discuss in the next paragraph, 
it is just the phase which shows the most striking change in the dynamics as the critical line is crossed.

\subsubsection*{Strong Quenches: $u_f>u_{fc}$}

As we anticipated, for quenches above the critical value $u_{fc}$ the dynamics of the system is qualitatively different,
reflecting the change in the behavior of the effective potential inversion points, see equation (\ref{eqn:Dplus}).
Let us start discussing the dynamics of double occupancy. Since the effective potential $\Gamma\left(D\right)$ 
has two simple roots, the motion of double occupation $D(t)$ is still periodic. 
However, the period $\m{T}$ and the amplitude $\m{A}$ of these strong coupling oscillations \emph{decrease} 
upon increasing the strength of the quench, in contrast to the weak quench case. Indeed the latter simply reads
$\m{A}=D_i-D_+\sim 1/u_f$ while the period reads
\be\label{eqn:period_strong}
\m{T} = \frac{4\,K\left(1/k\right)}{\sqrt{u_f\left(u_f-u_i\right)}}\,,
\ee
with argument $1/k$ given by
\be
\frac{1}{k}=\sqrt{\frac{2D_i\,u_{fc}}{u_f\left(u_f-u_i\right)}}\,.
\ee
Deep in the strong coupling regime, $u_f\gg\,u_i$, we get 
\be\label{eqn:period_sc}
\m{T}\simeq\,\frac{2\pi}{u_f}\,,
\ee
smoothly matching the atomic limit result. Hence the resulting dynamics shows very fast oscillations with a reduced
 amplitude. In the strong quench limit the double occupation dynamics is completely frozen, doublons have no available 
elastic channel to decay \cite{Rosch_prl08}.

As the critical quench line $u_{fc}$ is approached from above the period of oscillations
shows the same logarithmic singularity found on the weak-coupling side.
From equation (\ref{eqn:period_strong}) we immediately see that
\be
\m{T}\sim \frac{4}{\sqrt{1-u_i^2}}\,\log\left(\frac{1}{u_f-u_{fc}}\right)\,,
\ee
namely the same singularity, with the same prefactor, appears on the two side of the dynamical transition.

As already anticipated, it is interesting to discuss the dynamics of the phase $\phi(t)$ when the quench is above 
the critical line. As shown in figure~\ref{fig:fig4}, as soon as the critical 
line is crossed, the phase starts precessing around the whole 
circle $(0,2\pi)$. This transition from a localized phase 
with small oscillations around $\phi=0$ to a 
delocalized phase where the dynamics 
is unbounded is, from a mathematical point of view, completely analogous to what happen in a simple pendulum. 
Right at the critical quench line the dynamics is on the separatrix and the phase takes infinite time to reach its 
metastable configuration. As we are going to see in the next paragraph this metastable configuration corresponds to a 
featureless Mott Insulator.
Before concluding, let's briefly discuss the dynamics of quasiparticle weight $Z(t)$ in the strong quench regime.
As we see in figure \ref{fig:fig3}, similarly to the double occupation, also $Z(t)$ shows fast oscillations 
with a period $\m{T}$ given by (\ref{eqn:period_sc}) at strong coupling. 
Interestingly, the amplitude $\m{A}_Z$ of those oscillations goes 
all the way to zero and keeps finite even for very large $u_f$. This can be easily understood by looking at 
the dependence of the quasiparticle weight from the phase $\phi(t)$. At half filling this simply reads
(\ref{eqn:Z_qp_onesite})
\be
Z(t)=16\,D(t)\left(1/2-D(t)\right)\,\cos^2\phi(t)\,,
\ee
from which we can conclude that, although the double occupation is neither zero nor one half, the quasiparticle
weight can vanish due to its phase dependence. As a result of this vanishing minimum we conclude that for $u_f\gg1$,
even though the dynamics of double occupancy gets frozen in the initial state, 
the amplitude of oscillations for $Z$ remains constant and equal to $\m{A}_Z=1-u_i^2$.
 
\begin{figure}[ht]
\begin{center}
\epsfig{figure=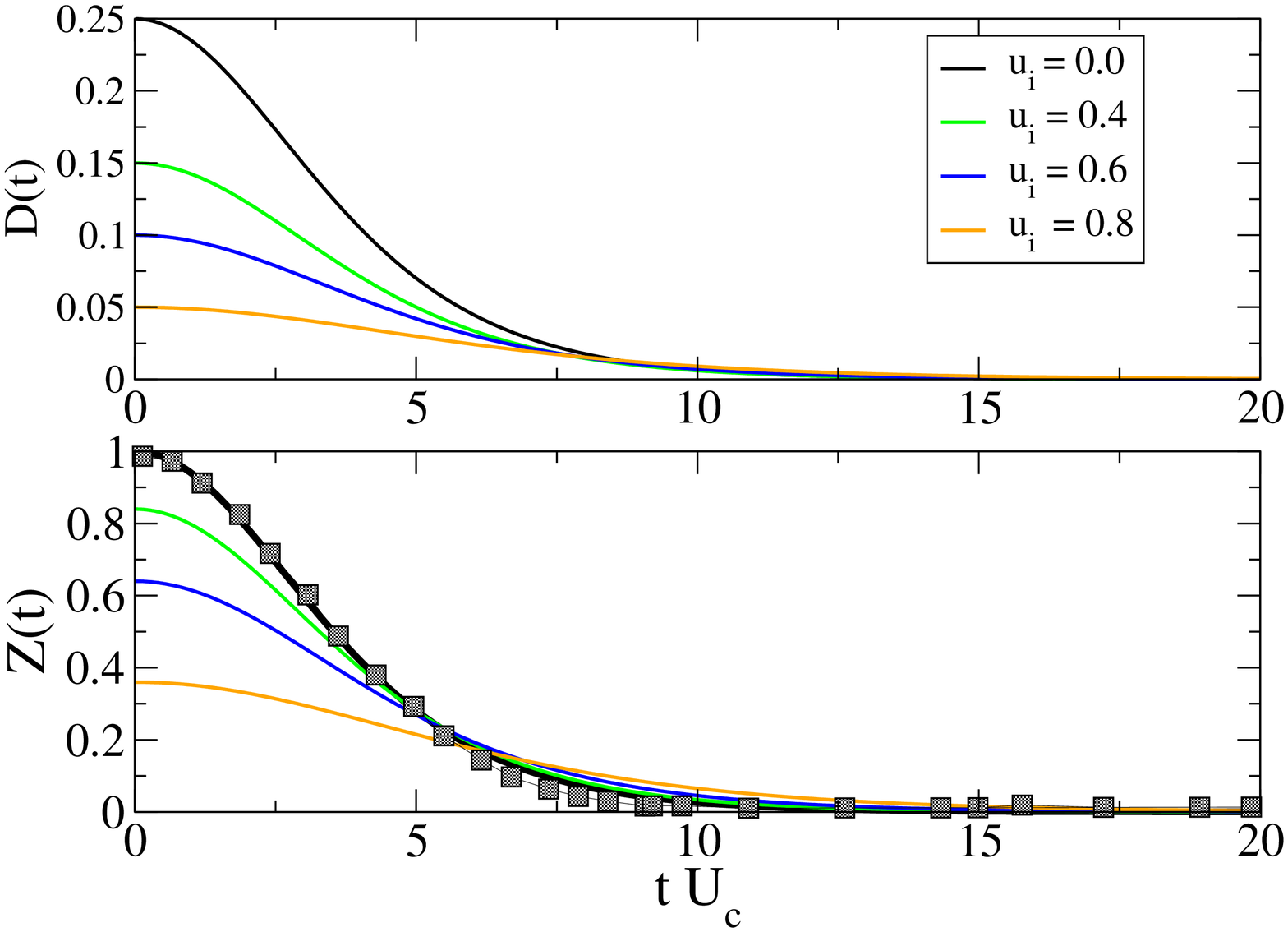,scale=0.325}
\caption{Dynamics after a quench at the critical interaction $u_{fc}$, for different initial conditions $u_i$.
Both double occupation $D(t)$ and quasiparticle weight $Z(t)$ decay exponentially to zero with a relaxation time 
$\tau_{\star}\sim1/\sqrt{Z_i}$ which increases with $u_i$ approaching the initial Mott Insulator. In the bottom panel
we compare the Gutzwiller results with those of DMFT (points, from Ref.~\cite{Werner_prl09}) for a quench starting from the non interacting limit.}
\label{fig:fig5}
\end{center}
\end{figure}

\subsubsection*{Critical Line}

Quite interestingly, the weak and the strong coupling regimes that we have so far discussed are separated by a critical
quench line $u_{fc}$ at which mean-field dynamics exhibits \emph{exponential relaxation}. This can be seen explicitly 
since in this limit the effective potential is simply given by $\Gamma\left(D\right)=\,D\sqrt{2u_{fc}\left(D_i-D\right)}$, 
thus the dynamics can be easily integrated to obtain the double occupation $D(t)$ at the critical quench line,
\be
D(t)=D_i\left(1-\tanh^2\left(t/\tau_{\star}\right)\right)\,.
\ee
We notice that, independently on the initial value of the correlation $u_i$, for $u_f=u_{fc}$ the double occupancy relaxes 
toward \emph{zero} with a characteristic time scale $\tau_{\star}=4/\sqrt{Z_i}$ that increases upon approaching the Mott insulator
$u_i\rightarrow 1$.
Analogously, also the quasiparticle weight $Z(t)$ approaches zero for long-time, with the same 
exponential behavior,
$$
Z(t)=Z_i\left(1-\tanh^2\left(t/\tau_{\star}\right)\right)\,.
$$
Since this is the only case in which our mean field dynamics
features a long-time steady state it is worth to compare the above behavior to the DMFT results \cite{Werner_prl09,Werner_long10}. 
In figure \ref{fig:fig5} we plot the behavior of the quasiparticle residue $Z(t)$ in the two approaches for the case $u_i=0$.
As we see they both vanishes at long times with a quite good agreement on the time scale. A similar comparison cannot
be done for the double occupation $D(t)$ which vanishes at long times in our mean field theory while saturates to a finite 
small value in DMFT. This is not surprising but again reflects the fact that our mean field dynamics cannot capture
the role of incoherent excitations. The long time vanishing of the quasiparticle weight has been interpreted in 
Refs.~\onlinecite{Werner_prl09,Werner_long10} 
as a signature of thermalization. Although we cannot comment on this issue, since our mean field theory cannot account
for thermalization, it is interesting to add some considerations.
From our results we see that for quenches at the critical line $u_{fc}$ the system reaches a steady state featuring
a complete suppression of charge fluctuations, namely $D=0$ and $Z=0$. This suggests that the above critical line $u_{fc}$ 
is obtained by tuning the initial energy $E_0$ of the quenched correlated metal to the energy of a collection of
decoupled half filled sites, the \emph{ideal} $t_{ij}=0$ Mott insulator. Indeed from this condition we immediately get
\be
E_0\left(u_{fc},u_i\right) = E_{Mott}\longrightarrow u_{fc}=\frac{1+u_i}{2}\,.
\ee
Surprisingly enough we find that the above condition gives a remarkable good agreement for the dynamical critical point found
in DMFT. 
Indeed if we use that latter criterium, we find an estimate for the 
critical $U_{fc}$ in units of the hopping integral $t$ and strating from 
$U_i=0$:
\be\label{eqn:dmft}
U_{fc} = 4\left|E_{kin}\right| \simeq 3.3 \,,
\ee
where $E_{kin}$ is the energy of a half-filled Fermi sea with a  
semielliptic density of states. Eq.~\eqn{eqn:dmft} is surprisingly close to 
the result of Refs.~\onlinecite{Werner_prl09,Werner_long10}.

\subsection{Long-time Averages}\label{sect:longtime}

As we have seen so far, the mean field Gutzwiller dynamics is periodic in the main part of the phase diagram excluding
the quench to the critical value $u_{fc}$ where an exponential behavior emerges. In spite of that, 
it is however worth to investigate a properly defined \emph{long-time} behavior of the dynamics which, as we are going
to show, features many interesting properties. To this extent we firstly introduce, for any given function $O(t)$ an 
\emph{integrated} (average) dynamics defined through
\be\label{eqn:average_dyn}
\langle O\rangle_t = \frac{1}{t}\,\int_0^t\,dt'\,O\left(t'\right)\,.
\ee
Then it is natural to define the long-time average as
\be\label{eqn:long_time_aver}
\bar{O}=\mbox{lim}_{t\rightarrow\infty}\,\langle O\rangle_t\,.
\ee
Notice that, since the relevant observables are periodic functions of time with period $\m{T}_O$ admitting a Fourier 
decomposition
the above definition~(\ref{eqn:long_time_aver}) can be equivalently written as
\be\label{eqn:aver}
\bar{O}=\frac{1}{\m{T}_O}\,\int_{\m{T}_O}\,dt\,O(t)\,.
\ee

We now study the behavior of steady state averages as a function of the initial and final values of 
the interaction. We consider the half filled case and for simplicity we assume $u_f>u_i$. 
Using equation (\ref{eqn:aver}) the average double occupation $\bar{D}$ can be written as
which reads
\be\label{eqn:D_aver}
\bar{D}=\frac{2}{\m{T}}\,\int_{D_{+}}^{D_i}\,\frac{D\,dD}{\sqrt{\Gamma\left(D\right)}}\,,
\ee
where $D_i$ and $D_+$ has been defined in the previous section. 
In addition, due to energy conservation, the knowledge of the average double occupancy $\bar{D}$ completely fixes
the average quasiparticle weight which reads 
\be\label{eqn:Z_aver}
\bar{Z}=Z_i+8u_f\left(\bar{D}-D_i\right)\,.
\ee

We now evaluate the long-time average $\bar{D}$ and $\bar{Z}$ as given in Eq. (\ref{eqn:D_aver}-\ref{eqn:Z_aver}) in 
the two different dynamical regimes we have previously identified.

\subsubsection*{Weak Quenches: $u_f<u_{fc}$}

\begin{figure}[t]
\begin{center}
\epsfig{figure=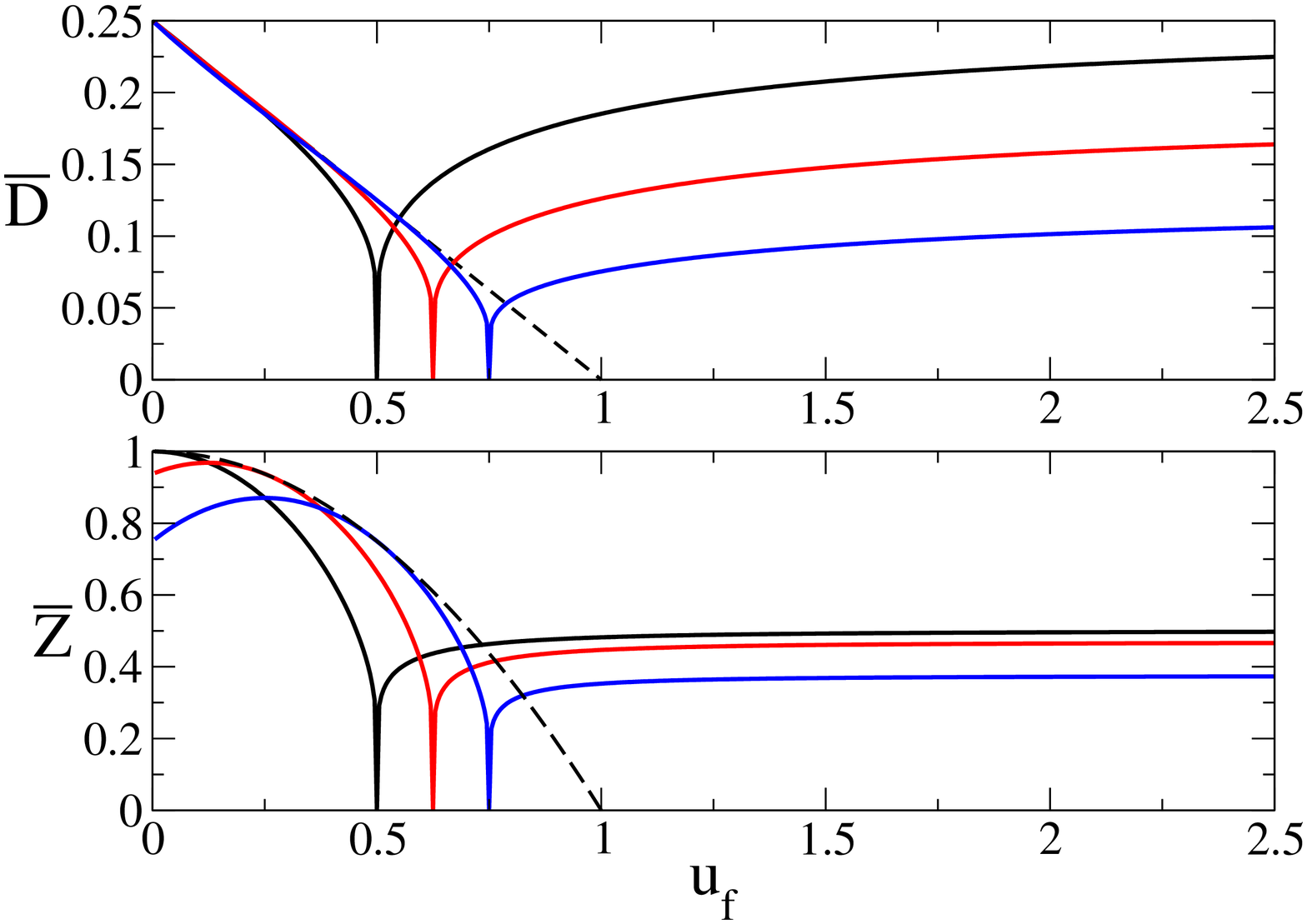,scale=0.325}
\caption{Average double occupation $\bar{D}$ (top) and quasiparticle weight $\bar{Z}$ (bottom) as a function of $u_f$ at
 fixed $u_i=0.0,0.25,0.5$ compared to the zero temperature equilibrium result (dashed lines).}
\label{fig:fig6}
\end{center}
\end{figure}

In the weak coupling regime and for $u_f>u_i$ the average double occupation at long times reads
\bea
\bar{D} &= &D_i\left(1-\frac{u_{fc}}{u_f}\right)+D_i\frac{u_{fc}}{u_f}\,\frac{E\left(k\right)}{K\left(k\right)}=\nonumber\\
&=&D_i\left[1+\frac{u_{fc}}{u_f}\left(\frac{E(k)-K(k)}{K(k)}\right)\right]\,,
\eea
where $K(k)$ and $E(k)$ are, respectively, the complete elliptic integrals of the first and second kind with argument
$k^2=\frac{u_f\left(u_f-u_i\right)}{2D_i\,u_{fc}}$. Similarly using the Eq. (\ref{eqn:Z_aver}) we get for the average
 quasiparticle weight the result
\be
\bar{Z}=Z_i\,\frac{E(k)}{K(k)}\,.
\ee
It is interesting to consider the asymptotic regime of a small quantum quench $\delta u =u_f-u_i\rightarrow0$. 
Then we can expand the elliptic integrals for small $k$ to get
\be
\bar{D} \simeq D_i -\frac{\delta u}{4}=\frac{1-u_f}{4}\,.
\ee
We see therefore that for small quenches the double occupation follows the zero temperature equilibrium curve, 
independently on the initial value of the interaction $u_i$. This is clearly shown in figure \ref{fig:fig6}. 
Since to lowest order in $\delta u$ no heating effects arise, this result implies that after a small quench of the
 interaction the average double occupation $\bar{D}$ is thermalized.
In addition, this result has an interesting consequence for what concerns the behavior of the quasiparticle weight $\bar{Z}$. A simple calculation to lowest order in $\delta u$ gives 
\be\label{eqn:Z_weak}
\bar{Z}\simeq Z_i-2\,u_f\,\left(u_f-u_i\right)\,,
\ee
from which we conclude that, as opposite to the double occupation $\bar{D}$, the long-time average quasiparticle weight 
differs from the zero temperature equilibrium result even at lowest order in the quench $\delta u$. In particular 
if we evaluate $\bar{Z}$ for the special case of a quench from a non interacting Fermi Sea ($u_i=0$) for which $Z_i=1$
we get the result, 
\be
1-\bar{Z}\left(u_f\right)=2\left(1-Z_{eq}\left(u_f\right)\right)\,,
\ee
firstly obtained in Ref.~\onlinecite{Kehrein_prl08} within the flow equation approach.  This peculiar mismatch between the zero 
temperature equilibrium quasiparticle residue and its non equilibrium counterpart is a general result of quenching a 
Fermi Sea \cite{Moeckel2009,MoeckelPhDThesis}. It signals the onset of a \emph{prethermal} regime where quasiparticle
 are well defined objects, momentum-averaged quantities such as kinetic and potential energy are thermalized while 
relaxation of distribution functions is delayed to later time scales. 
 We note that our simple mean field 
theory correctly captures the onset of this long-lived state but fails in 
describing its subsequent relaxation towards  
equilibrium. 

Interestingly, when approaching the critical quench line from below the average double occupation $\bar{D}$ vanishes 
logarithmically. Indeed for $k\rightarrow1$ we have
\be
K(k)\simeq \mbox{log}\left(4/\sqrt{1-k^2}\right) + O\left(1-k^2\right)\,,
\ee
and 
\be
E\left(k\right)\simeq 1 + O\left(1-k^2\right)\,,
\ee
therefore
\be
\bar{D}\simeq D_i\left(\frac{u_f-u_{fc}}{u_f}\right) +\frac{2D_i}{\mbox{log}\left(\frac{1}{u_{fc}-u_f}\right)}\,,
\ee
The leading term is therefore logarithmic as mentioned, with linear corrections in $\delta u =u_{fc}-u_f$
\be
\bar{D}\simeq\frac{2D_i}{\mbox{log}\left(\frac{1}{u_{fc}-u_f}\right)}
\left(1+\frac{\delta u\,\mbox{log}\,\delta u}{2u_{fc}}\right)\,,
\ee
A similar behavior is found for the quasiparticle weight $\bar{Z}$ which reads 
\be
\bar{Z}\simeq\frac{2Z_i}{\mbox{log}\left(\frac{1}{u_{fc}-u_f}\right)}\,,
\ee

\subsubsection*{Strong Quenches: $u_f>u_{fc}$}

In the strong coupling regime the average double occupation reads
\be
\bar{D} = \frac{u_{fc}-u_f}{2} + \frac{u_f-u_i}{2}\,\frac{E\left(k\right)}{K\left(k\right)}\,,
\ee
with the argument given by $k^2=\frac{2\,D_i\,u_{fc}}{u_f\left(u_f-u_i\right)}$.

Deep in the strong coupling regime, $u_f\gg u_i$, $k$ goes to zero and we can use the asymptotic for $E(k)$ and $K(k)$
\be
\frac{E\left(k\right)}{K\left(k\right)}\simeq 1-\frac{k^2}{2}\, ,
\ee
to obtain
\be\label{eqn:docc_strong}
\bar{D}\simeq D_i\left(1 -\frac{u_{fc}}{2u_f}\right).
\ee
We see therefore that, for infinitely large quenches, $u_f\rightarrow\infty$, the dynamics is trapped into the initial state.
Interestingly enough, for quenches starting from $u_i=0$ the scaling 
\eqn{eqn:docc_strong} exactly matches the strong 
coupling perturbative result obtained in 
Ref.~\onlinecite{Werner_prl09} for the prethermal plateau.
Indeed using the fact that for $u_i=0$ we have $D_i\,u_{fc}=1/8=\vert\bar{\eps}\vert$, where $\bar{\eps}$ is the kinetic 
energy of the half-filled Fermi Sea, we find
$$
\bar{D}\simeq D_i-\frac{\vert\bar{\eps}\vert}{2U_f}\,,
$$
in accordance with strong coupling perturbation theory.
The agreement at strong coupling is remarkable if thought from the point of view of thermal 
equilibrium, where one knows the Gutzwiller wavefunction cannot capture the Hubbard bands,
and suggests that our Gutzwiller ansatz can interpolate between the weak and the strong coupling dynamical regime. 

As opposite, when approaching the critical quench line from above we obtain a vanishing long-time average, with the same 
logarithmic behavior we have found on the weak coupling side. Indeed for $u_f\rightarrow\,u_{fc}$ from above we have that 
$k\rightarrow 1^-$ and therefore we can again make use of the asymptotic for the complete elliptic integrals. 
We thus obtain
\be
\bar{D}\simeq\frac{2D_i}{\mbox{log}\left(\frac{1}{u_f-u_{fc}}\right)}
\left(1+\frac{\delta u\,\mbox{log}\,\delta u}{4D_i}\right)\,.
\ee
Note that the approach to zero is the same in both sides of the phase diagram, while the corrections are slightly different.

For what concerns the quasiparticle weight $\bar{Z}$ to get the leading behavior $o\left(1/u_f\right)$ we need the 
double occupancy to next-to-leading order. Expanding the ratio between elliptic functions we get
\be
\frac{E\left(k\right)}{K\left(k\right)}\simeq 1-\frac{k^2}{2}-\frac{k^4}{8}+
O\left(k^6\right)\,,
\ee
and using the expression for $k\simeq Z_i/4u_f^2$ we obtain the following asymptotic behavior for $\bar{Z}$
\be
\bar{Z}\simeq Z_i -2u_f^2\,k^2\left(1+k^2/4\right)\simeq
\frac{Z_i}{2}\left(1-\frac{Z_i}{16u_f^2}\right)\,,
\ee
which shows that also $\bar{Z}$ increases from the critical line to large $u_f$ and deep in the strong coupling regime 
it saturates to a finite plateau which, however does not coincide with its initial value $Z_i$ but rather it is smaller 
by a factor of two due to energy conservation.

\subsection{Quench Dynamics away from half-filling}

An important outcome of previous sections has been the identification of a critical interaction quench $u_{fc}$ 
where an exponentially fast relaxation emerges. This value of quenches separates two different dynamical regimes where the 
system gets trapped into metastable prethermal states. In order to understand the origin of such a sharp transition and its
possible relation to equilibrium critical point of the Hubbard model it is natural to extend the mean field analysis away from half-filling, 
where no transition between a Metal and a Mott Insulator exists in equilibrium.
This can be done straightforwardly, for example, by a direct integration of the mean field equations of motion
(\ref{eqn:pendulum1}-\ref{eqn:pendulum2}). It is however more instructive to 
proceed again by considering the effective dynamics for the double occupation, obtained using the conservation of energy, 
that we wrote as
\be
\dot{D}=\sqrt{\Gamma(D)}\,,
\ee
We now argue that any finite doping $\delta$ is enough to wash out the dynamical critical point and turn it into a crossover.
To see this, it is worth to consider again the effective potential $\Gamma\left(D\right)$ which enters the 
above dynamics. Indeed the qualitative analysis we have performed in section 
\ref{sect:half_filling} can be done even for finite doping $\delta$. 
As we will show explicitly in the Appendix \ref{app:doping}
the effective potential keep the same structure for $\delta\neq0$, with three inversion points respectively given by $D_i$ 
- the zero temperature finite doping Gutzwiller solution - and $D_{\pm}$. 
\begin{figure}[t]
\begin{center}
\epsfig{figure=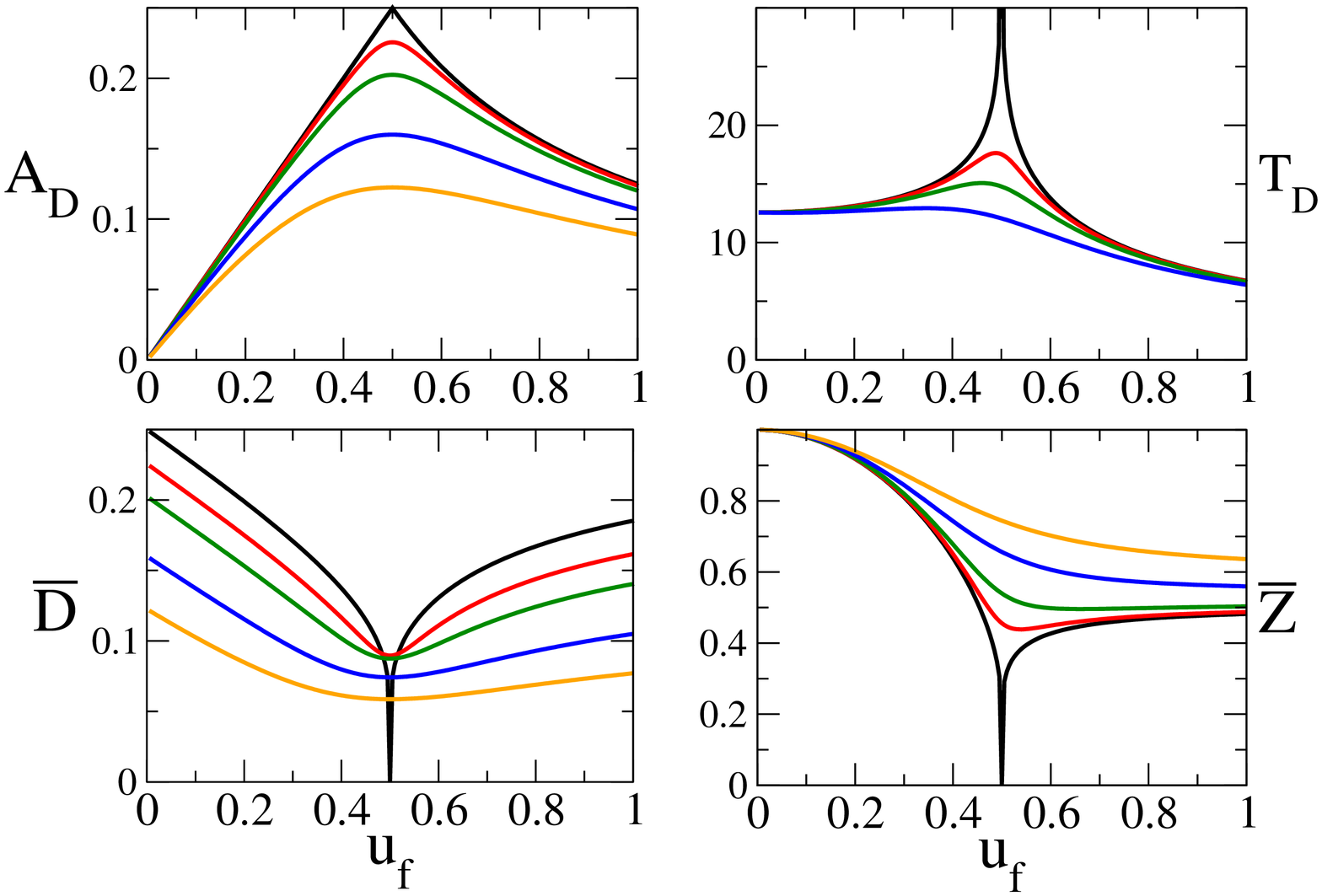,scale=0.325}
\caption{Top Panel: amplitude $\m{A}_D$ (left) and period $\m{T}_D$ at $u_i=0$ and  $\delta=0.0,0.05,0.10,0.20,0.30$. 
Bottom Panel: averages double occupancy $\bar{D}$ (left) and quasiparticle weight $\bar{Z}$ (right) at $u_i=0$ and
 $\delta=0.0,0.05,0.10,0.20,0.30$.}
\label{fig:fig7}
\end{center}
\end{figure}
As a consequence, all the differences between the doped and the half-filled case are hidden in the behavior of 
the two non-trivial roots $D_{+},D_-$ as a function of $u_f$. 
Their explicit expression is quite lengthy and it is reported for completeness in Appendix \ref{app:doping}. 
As we can see from figure \ref{fig:fig2}, 
those two roots, which at half-filling are degenerate at $u_{fc}$,
are always distinct at finite doping. In particular 
at the half-filling critical quench line $u_{fc}$,  we find at finite doping  
$$
D_{+}(u_{fc})-D_-(u_{fc}) \simeq \delta\,.
$$
As a consequence the dynamics of double occupancy (and hence of quasiparticle weight) always features a \emph{finite}
period given by
\be
\m{T}\simeq\frac{K\left(k\right)}{\sqrt{u_f\left(D_i-D_-\right)}}\,,
\ee
with the argument $k$ of the elliptic function defined in term of the inversion points as
\be
k=\sqrt{\left(D_i-D_+\right)/\left(D_i-D_-\right)}\,.
\ee
Notice that, since the two inversion points never collapse $D_+ > D_-$, the argument $k$ is always strictly
lesser than one, $k<1$, and no singularity in $\m{T}$ arises.

 In figure \ref{fig:fig7} (top panels) we plot the period
$\m{T}$ and the amplitude $\m{A}$ of the double occupancy oscillations in the doped case, as a function of $u_f$ at 
fixed $u_{i}$. We notice that both quantities are smooth across $u_{fc}$, and in particular the logarithmic singularity
 in the period turns into a sharp peak which broadens out as the doping increases. 

We finally remark that a small doping not only affects the dynamics, but also drastically changes the long-time averages 
properties with respect to the results we have depicted in section \ref{sect:half_filling}. This can be worked out 
explicitly by using the same equations we have obtained for the half-filling case, cfr. section \ref{sect:longtime},
 provided the correct expression for the roots $D_{+},D_-$ is used. As we can see from figure \ref{fig:fig7} both double 
occupation and quasiparticle weight stay always finite as $u_f$ increases and only show a dip around the critical quench 
line which is gradually smoothed out as the doping increases.

In conclusion, we have shown that the dynamical transition described in section \ref{sect:half_filling} is a peculiar feature 
of the half-filled case, namely that any finite doping $\delta\neq0$ is enough to wash out this dynamical 
transition, cutting off the logarithmic divergence in the oscillation period $\m{T}$.

\subsection{Discussion}

We conclude this section by discussing the results of our time dependent mean field theory for the fermionic Hubbard model 
in light of those recently obtained in the literature using different approaches, such as the Flow Equation method~
\cite{Kehrein_prl08,Moeckel2009} and the Non Equilibrium Dynamical Mean Field Theory~\cite{Werner_prl09,Werner_long10},
both of which considered a quantum quench starting from an half-filled non interacting Fermi Sea.
As we already mentioned, our mean field results feature an oversimplified periodical dynamics that lacks relaxation 
to a steady state at long times. This can be traced back to the suppression of quantum fluctuations which is at the
ground of our treatment. In this respect we notice that both approaches work much better, displayng some 
damping at long times. Beside this obvious drawback we can say that, quite remarkably, a mean field theory catches many 
interesting features of the problem.  

First of all our variational ansatz is able to capture both regimes of pre-thermalization found at weak
\cite{Kehrein_prl08} and strong coupling \cite{Werner_prl09}. Those long lived metastable regimes, which are, 
respectively, due to Fermi statistics  and to long-lived double occupations,  are quantitatively reproduced by our approach 
as it appears clearly from the analysis of long time averages 
(see Eqs.~\eqn{eqn:Z_weak} and \eqn{eqn:docc_strong}).
However, as generally expected in mean field theories, those metastable states are  wrongly predicted to have  
infinite lifetime. 
A second interesting point that clearly emerges from our analysis is 
the existence of a dynamical critical line that separates 
those two distinct regimes, and where an exponentially fast 
relaxation emerges, as firstly shown in Ref.~\onlinecite{Werner_prl09}.
On one hand, the existence of a dynamical critical point could be anticipated 
since at equilibrium the model undergoes a quantum phase transition, the 
Mott transition. Indeed, as we have shown, any finite doping 
turns the dynamical transition into a crossover. On the other hand, 
it was noted in Ref.~\onlinecite{Werner_prl09} that the energy pumped in the 
quench at $U_{fc}$ with $U_i=0$ would correspond, should thermalization be 
assumed, to an effective temperature $T_\star$ higher  
than the Mott ending point, where no critical dynamics could have been 
foreseen. Such an observation points to a dynamical transition that could be 
associated with loss of ergodicity and which is not incompatible with our    
finding that the critical quench occurs when the correlated metal 
is initially prepared with the energy
of the \emph{ideal} Mott insulator, a collection of independent sites. 
Interestingly enough, 
such a condition gives a excellent match with the DMFT estimate
of the dynamical critical point (see equation \ref{eqn:dmft}).
Finally, we note that the issue of a non equilibrium dynamical transition 
in the quench dynamics of interacting quantum 
systems seems to be of more general interest. Indeed recent investigations on the fully 
connected Bose Hubbard model~\cite{SciollaBiroli_prl10} and the scalar $\phi^4$ mean field theory
~\cite{GambassiCalabrese_preprint} reveals that a very similar phenomenon is present in 
these models as well. Whether this is an artifact of the mean field approximation or rather a generic feature 
of the quench dynamics of interacting quantum systems in more than one dimension, as recent works would suggest,
\cite{GambassiCalabrese_preprint} is an interesting subject that requires further investigations. 
In this respect an interesting question is the role played by small quantum fluctuations on such a dynamical transition. 
We will try to partially address this issue in the remaining part of this paper.

\section{$Z_2$ Slave Spin Formulation}\label{sect:Ising}

We have shown that, within the Gutzwiller approximation, the variational 
principle when applied to the Shr{\oe}dinger equation amounts to determine the 
saddle point of an action $S\left[\phi_{i\alpha},O_{i\alpha},\Phi\right]$ that 
depends on pairs of conjugate fields, $\phi_{i\alpha}$ and $O_{i\alpha}$, and 
on a Slater determinant or BCS wavefunction. The saddle point reduces to a set 
of first order coupled differential equations for the conjugate fields and 
for the average values of single particle operators on $|\Phi(t)\rangle$. 
One could be tempted to interpret this result as the mean field decoupling 
of the Heisenberg equations of motion for the average values of 
a set of quantum operators corresponding to some effective quantum Hamiltonian. 
Identifying such a quantum Hamiltonian could then allow adding quantum 
fluctuations on top of the mean field results. This is right the same 
conceptual scheme invoked to associate the time-dependent Hartree-Fock 
equations to an effective Hamiltonian of non-interacting bosons that represent 
particle-hole excitations. In our case we would expect the quantum Hamiltonian 
to describe free electrons coupled to a set of conjugate Bose fields,    
$\phi_{i\alpha}$ and $O_{i\alpha}$, which in fact resembles the 
conventional slave-boson approaches to correlated systems. 
 
We are going to show that this program can be easily accomplished in the simple 
Hubbard model, although in a different and more rigorous manner 
than simply quantizing the classical equations of motion. 
To this extent we formulate the original Hubbard model 
in terms of an auxiliary Quantum Ising Model in a transverse field coupled to free fermionic quasiparticles, 
in the framework of the recently
introduced $Z_2$ slave spin theory~\cite{Z2-1,Z2-2}.

\subsection{Mapping onto a Quantum Ising Model in a Transverse Field}

The idea of writing the Hubbard model in terms of auxiliary spins coupled to free quasiparticles 
is not new.\cite{DeMedici-1,DeMedici-2}
A minimal formulation in terms of a single Ising spin and a fermionic degrees of freedom 
has been recently introduced,\cite{Z2-1,Z2-2}
based on a mapping between the local physical Hilbert space of the Hubbard Model and the Hilbert space of the 
auxiliary model subjected to a constraint. Here we derive the same mapping by showing that the 
identification holds for the partition functions as well, when evaluated order by order 
in perturbation theory in $U$. The advantage of this alternative formulation is that the role of 
the lattice coordination emerges more clearly. 

We write the Hubbard interaction as  
\[
U\,n_\up\,n_\down = \frac{U}{4}\,\Big[2(n-1)^2-1\Big] + 
\frac{U}{4}\,(2n-1).
\]
The last term can be absorbed into the chemical potential, so that we 
shall consider as interaction only the first term. We define  
\be
2(n-1)^2-1 = \mathrm{e}^{i\pi\,n}\equiv \Omega,
\label{1}
\ee
where the operator $\Omega$ is real and unitary and has 
eigenvalues $-1$ for $n=1$ and $+1$ for $n=0,2$. It follows that 
\be
\Omega\,c^\dagger_\sigma\,\Omega = -c^\dagger_\sigma,
\label{2}
\ee
namely it changes sign to the fermion operator. 

Let us concentrate on a given site, with local energy $\epsilon$,  
whose Fermi operator we shall denote as $c^\dagger_\sigma$ 
and density operator $n$. 
The rest of the lattice sites, Fermi operators $d_{\bR\sigma}$, are 
described by the generically interacting Hamiltonian $H_{bath}$ and are coupled 
to the site under investigation by 
\be
H_{tunn} = -\sum_{\bR\sigma}\,t_\bR\,c^\dagger_\sigma d^\dagga_{\bR\sigma} + H.c..
\ee
We shall denote as 
\[
H_0 = H_{bath} + \epsilon\,n +H_{tunn},
\]
the {\sl unperturbed} Hamiltonian and
\[
\fract{U}{4}\,\mathrm{e}^{i\pi\,n} = \fract{U}{4}\,\Omega,
\]
the perturbation. 
Suppose we calculate the partition function within perturbation theory. 
A generic $n$-th order correction to the partition function is
\begin{widetext}
\ba
Z^{(n)} &=& \left(-\fract{U}{4}\right)^n\,\int_0^\beta\,d\tau_1\,\int_0^{\tau_1}\,d\tau_2
\, \dots \int_0^{\tau_{n-1}}\, d\tau_n\,\Tr\Bigg[ 
\mathrm{e}^{-\left(\beta-\tau_1\right)\,H_0}\,\Omega\, 
\mathrm{e}^{-\left(\tau_1-\tau_2\right)\,H_0}\,\Omega\,\dots\, 
\Omega\,\mathrm{e}^{-\left(\tau_{n-1}-\tau_n\right)\,H_0}\,\Omega\,
\mathrm{e}^{-\left(\tau_n-0\right)\,H_0}\Bigg]\,,
\ea
\end{widetext}
Because of \eqn{2} $\Omega\,H_0\,\Omega = H_{bath} + \epsilon\,n -H_{tunn} \equiv H_1$. 
We shall distinguish the two cases of $n$ even or odd. 
In the even case one easily realizes that
\begin{widetext}
\bea
Z^{(2n)} &=& \left(\fract{U}{4}\right)^{2n}\,\int_0^\beta\,d\tau_1\,\int_0^{\tau_1}\,d\tau_2
\, \dots \int_0^{\tau_{2n-1}}\, d\tau_{2n}\,\nonumber \\
&& ~~~~~~~~ \Tr\Bigg[ 
\mathrm{e}^{-\left(\beta-\tau_1\right)\,H_0}\, 
\mathrm{e}^{-\left(\tau_1-\tau_2\right)\,H_1}\,\mathrm{e}^{-\left(\tau_2-\tau_3\right)\,H_0}\,
\dots\, 
\mathrm{e}^{-\left(\tau_{2n-1}-\tau_{2n}\right)\,H_1}\,
\mathrm{e}^{-\left(\tau_{2n}-0\right)\,H_0}\Bigg],\label{n-even}
\eea  
\end{widetext}
which resembles an iterated X-ray edge problem, like 
in the Anderson-Yuval representation 
of the Kondo model. We note that, since $\Omega^2=1$, Eq.~\eqn{n-even} is invariant under 
$H_0 \leftrightarrow H_1$. For the odd case, one finds instead
\begin{widetext}
\bea
&&Z^{(2n+1)} = -\left(\fract{U}{4}\right)^{2n+1}\,\int_0^\beta\,d\tau_1\,\int_0^{\tau_1}\,d\tau_2
\, \dots \int_0^{\tau_{2n}}\, d\tau_{2n+1}\,\nonumber \\
&& \Tr\Bigg[ 
\mathrm{e}^{-\left(\beta-\tau_1\right)\,H_0}\,
\mathrm{e}^{-\left(\tau_1-\tau_2\right)\,H_1}\,\mathrm{e}^{-\left(\tau_2-\tau_3\right)\,H_0}\,
\dots\, 
\mathrm{e}^{-\left(\tau_{2n-1}-\tau_{2n}\right)\,H_1}\,\mathrm{e}^{-\left(\tau_{2n}-\tau_{2n+1}\right)\,H_0}\,
\mathrm{e}^{-\left(\tau_{2n+1}-0\right)\,H_1}\,\Omega\Bigg].\nonumber\\
&& \label{n-odd}
\eea 
\end{widetext}
Once again the above expression is also equal to that one where $H_0$ is interchanged with $H_1$.  

Can one reproduce the same perturbative expansion with some other model?
Let us consider an Ising-like Hamiltonian $H_{Ising}=H_*+H_{transv}$ where the 
unperturbed term is 
\be
H_* = H_{bath} + \epsilon\,n + \sigma^x\,H_{tunn},
\ee
the perturbation is
\be
H_{transv} = -\fract{U}{4}\,\sigma^z,\label{transverse}
\ee
and $\sigma^a$, $a=x,y,z$, are Pauli matrices.  
If we take the trace over eigenstates of $\sigma^x$ 
-- note that for $\sigma^x=1$ 
$H_*=H_0$, while for $\sigma^x=-1$ $H_*=H_1$ -- 
the perturbation \eqn{transverse} 
may act only an even number of times and one easily find that the final result 
is just twice \eqn{n-even}. In other words, $Z^{(2n)}$ is 
half of the $2n$-th order term in the 
perturbative expansion of the Ising model $H_{Ising}$. How do we get 
the odd order terms in the expansion? 
Let us consider the perturbative expansion of 
\[
-\Tr\Bigg(\mathrm{e}^{-\beta\,H_{Ising}}\,\sigma^z\,\Omega\Bigg).
\]
It is clear that now only odd terms in the expansion over 
eigenstates of $\sigma^x$ 
will contribute and one easily realizes that the final result 
is twice \eqn{n-odd}. 

Therefore, the partition function of the original model is also equal to 
\be
Z = \Tr\Bigg[\mathrm{e}^{-\beta\,H_{Ising}}\, 
\left(\fract{1-\sigma^z\,\Omega}{2}\right)\Bigg].
\label{map-1-intermedio}
\ee
We note that 
\be
\mathcal{Q} = \fract{1-\sigma^z\,\Omega}{2},\label{Z2-constraint}
\ee
is actually a projector of the enlarged Hilbert space onto the subspace where 
if $n=1$ then $\sigma^z=+1$ while, if $n=0,2$, then $\sigma^z=-1$.
As a matter of fact, $\mathcal{Q}$ is just the constraint introduced in 
Ref.~\onlinecite{Z2-2} as a basis of the $Z2$ slave-spin 
representation of the Hubbard model. In fact, what we have done here is 
simply re-deriving the mapping of Ref.~\onlinecite{Z2-2} in a different way.
There are however some interesting aspects of the mapping that emerge clearly 
at the level of the partition functions and were not discussed in   
Ref.~\onlinecite{Z2-2}.    

We note that what we have shown so far is that, given  
an Anderson impurity model with Hamiltonian 
\bea
H_{AIM} &=& H_{bath}  +  H_{tunn} + \epsilon\,n
+ \frac{U}{2}\,(n-1)^2 \nonumber\\ 
&=& H_{bath} + H_{tunn} + \epsilon\,n + 
\frac{U}{4}\,\left(1+\Omega\right),\label{H-AIM}
\eea
its partition function can be also written as 
\bea
Z_{AIM} &=& \frac{1}{2}\,\Tr\Bigg[\mathrm{e}^{-\beta\,H_{Ising}}\, 
\bigg(1-\sigma^z\,\Omega\bigg)\Bigg]\nonumber \\
&=& \frac{1}{2}\,Z_{Ising}\,\bigg(1-\langle \sigma^z\,\Omega\rangle\bigg),
\label{Z-AIM}
\eea
where 
\be
H_{Ising} = H_{bath} + \epsilon\,n + \sigma^x\,H_{tunn} + 
\frac{U}{4}\,\left(1-\sigma^z\right),
\label{H-Ising-AIM}
\ee
and 
\[
Z_{Ising}= \Tr\left(\mathrm{e}^{-\beta H_{Ising}}\right).
\]
As mentioned above, $Z_{Ising}$ is even in $U$, 
while $\langle \sigma^z\,\Omega\rangle$ is 
odd. As a simple byproduct, we note that, if particle-hole symmetry holds, the 
partition function must be even in $U$, so that 
\be
Z_{AIM} \equiv \frac{1}{2}\,Z_{Ising},\label{mapping-exact-AIM}
\ee
hence the constraint is uneffective and the mapping holds trivially.    
It was noticed in Ref.~\onlinecite{Z2-2} that $H_{Ising}$ in \eqn{H-Ising-AIM} 
possesses a local $Z_2$ gauge symmetry,  
$c^\dagger_\sigma \to -c^\dagger_\sigma$ and $\sigma^x\to -\sigma^x$, which can not 
be broken. Indeed, the factor $1/2$ in \eqn{mapping-exact-AIM} avoids the 
consequent double counting.

One can straightforwardly extend the above procedure to a 
collection of interacting sites, 
hence to the Hubbard model, with the final result that 
\be
Z = \Tr\Bigg[\mathcal{Q}\; 
\mathrm{e}^{-\beta\,H_{Ising}}\Bigg],
\ee
where now
\be
H_{Ising} = -t\sum_{<\bR,\bRp>\, \sigma}\, \sigma^x_\bR\,\sigma^x_\bRp\,
c^\dagger_{\bR\sigma}c^\dagga_{\bRp\sigma}
+ \frac{U}{4}\sum_\bR\,\left(1-\sigma^z_\bR\right).\label{H-Ising-Hubbard}
\ee
and the constraint is 
\be
\mathcal{Q} = \prod_\bR\left(\fract{1-\sigma^z_\bR\Omega_\bR}{2}\right).
\label{Z2-constraint-Hubbard}
\ee
We note that, if $\tau\to -it$, the mapping still holds and shows 
that the time-evolution of the Hubbard model can be mapped onto 
the time evolution of $H_{Ising}$. In particular, since  
$\left[\mathcal{Q},H_{Ising}\right]=0$, the two evolutions 
are exactly the same on a state that satisfies the constraint. 

\subsection{Recovering the Gutzwiller approximation at equilibrium}
\label{Recovering the Gutzwiller approximation at equilibrium}

Let us now consider a lattice whose coordination 
tends to infinity in a such a way that the hopping energy per site 
remains well defined. In this limit, it is well known\cite{Review_DMFT_96} 
that the Hubbard model maps onto 
an Anderson impurity model self-consistently coupled to a conduction bath. 
We showed earlier that when particle-hole symmetry holds, 
the constraint is uneffective for the mapping of the Anderson impurity 
model to the Ising model. It follows that the same holds 
also for the Hubbard model, in which case
\be
Z_{Hubbard} = \left(\frac{1}{2}\right)^N\,Z_{Ising},\label{Z-map-Hubbard}
\ee
where $N$ is the number of sites. 

Therefore, in infinite coordination lattices and at particle-hole symmetry, 
we could calculate the partition function of the model
\be
H_{Ising} = -\fract{t}{\sqrt{z}}\sum_{<\bR,\bRp>\, \sigma}\, \sigma^x_\bR\,\sigma^x_\bRp\,
c^\dagger_{\bR\sigma}c^\dagga_{\bRp\sigma}
+ \frac{U}{4}\sum_\bR\,\left(1-\sigma^z_\bR\right), \label{H-Ising-Hubbard-2}
\ee
and obtain that of the Hubbard model through \eqn{Z-map-Hubbard}. The factor $z$ in 
\eqn{H-Ising-Hubbard-2} is the lattice coordination and must be sent to infinity at the 
end of the calculation.\cite{Review_DMFT_96}  
It turns out that the Gutzwiller approximation is nothing but the mean field 
decoupling of $H_{Ising}$, assuming a wavefunction product 
of an Ising part times a fermionic one. The degeneracy of the solution that 
derives from the local $Z_2$ gauge symmetry, $\sigma^x_\bR\to - \sigma^x_\bR$ 
and $c^\dagger_{\bR\sigma} \to - c^\dagger_{\bR\sigma}$ is canceled out by the $(1/2)^N$ factor 
in \eqn{Z-map-Hubbard}. 

To recover the Gutzwiller result for the Mott transition, let us consider a trial translationally-invariant 
wavefunction $|\Psi\rangle = |\Phi_{\sigma}\rangle\,|\Phi_{c}\rangle$, 
where $|\Phi_\sigma\rangle$ is a Ising-spin state and $|\Phi_c\rangle$ an electron one.  
If we define 
\[
-t\fract{1}{\sqrt{z}}\,
\sum_\sigma\,\langle \Phi_{c}|\,c^\dagger_{\bR\sigma}c^\dagga_{\bRp\sigma}+H.c.\,|\Phi_{c}\rangle 
= -\frac{2}{z}\,\varepsilon,
\]
where $-\varepsilon$ is the average hopping energy per site of $|\Phi_{c}\rangle$, 
then the average value per site of the Hamiltonian \eqn{H-Ising-Hubbard-2} is 
\[
E = \langle \Phi_{\sigma}|\, -\varepsilon\, \fract{2}{z}\sum_{<\bR,\bRp>} \sigma^x_\bR\,\sigma^x_\bRp  
+ \frac{U}{4}\sum_\bR\left(1-\sigma^z_\bR\right)\,|\Phi_{\sigma}\rangle, 
\]
i.e. the energy of an Ising model in a transverse field. 
We assume $|\Phi_{\sigma}\rangle = \mathcal{U}\,|\Phi_0\rangle$ where the unitary operator 
\be
\mathcal{U} = 
\exp\left(i\frac{\beta}{2}\sum_\bR\,\sigma^y_\bR\right),\label{trial-Ising}
\ee
so that $E$ becomes the average value on $|\Phi_0\rangle$ of the Hamiltonian 
\bea
H_* &=& \frac{U}{4}\sum_\bR\, 1 - \cos\beta\,\sigma^z_\bR -\sin\beta\,\sigma^x_\bR\nonumber\\
&& -\varepsilon\,\frac{2}{z}\sum_{<\bR,\bRp>}\,\Big(\cos\beta\, \sigma^x_\bR -\sin\beta\,
\sigma^z_\bRp\Big)\nonumber \\
&& \phantom{-\varepsilon\,\frac{2}{z}\sum_{<i,j>}}\;\Big(\cos\beta\, \sigma^x_\bRp -\sin\beta\,
\sigma^z_\bRp\Big).\label{H_*-equi-1}
\eea
We assume that $|\Phi_0\rangle$ is so close to the fully ferromagnetic state with all spins 
oriented along $x$ that we can set  
\bea
\sigma^x_\bR &\simeq& 1 - \left(x_\bR^2+p_\bR^2-1\right) \equiv 1-\Pi_\bR,\label{sigma^x}\\
\sigma^y_\bR &\simeq& -\sqrt{2}\,p_\bR,\label{sigma^y}\\
\sigma^z_\bR &\simeq& \sqrt{2}\,x_\bR,\label{sigma^z}
\eea
where $x_\bR$ and $p_\bR$ are conjugate variables. 
If we substitute the above expressions in \eqn{H_*-equi-1} and fix $\beta$ in such a way that all terms linear 
in $x_\bR$ vanish, we find 
\be
\sin\beta = \fract{U}{8\varepsilon},\label{GA-sol-1}
\ee 
for $U<8\varepsilon \equiv U_c$, while $\sin\beta=1$ otherwise. $U_c$ is the mean-field value of 
critical transverse field that separates the ordered phase from the disordered one in the Ising model. 
It also identifies the Mott transition in the original Hubbard model, and, in fact, the value of $U_c$ 
coincides with that of the Gutzwiller approximation. Because of the above choice of $\beta$, once we 
expand the Hamiltonian \eqn{H_*-equi-1} up to second order in $x_\bR$ and $p_\bR$ we find, apart from 
constant terms and in units of $U_c$,  
\be
H_* \simeq \frac{a}{2}\sum_i\Big(x_\bR^2+p_\bR^2\Big) - \frac{b}{2}\,\frac{2}{z}\sum_{<\bR,\bRp>}\,x_\bR\,x_\bRp,
\label{H_*-metal-insulator}
\ee
where $a=1/2$ and $b=u^2/2$ for $u=(U/U_c)<1$, the metallic phase, while $a=u/2$ and $b=1/2$ for $u>1$, 
the Mott insulator. The spectrum of the excitations on both side of the transition is that of 
acoustic modes with dispersion in momentum space 
\be
\omega_{\mathbf{q}} = \sqrt{a\left(a-b\gamma_{\mathbf{q}}\right)},\label{omega_q}
\ee
where, assuming a hypercubic lattice in $d=z/2$ dimensions, 
\be
\gamma_{\mathbf{q}} = \frac{1}{d}\sum_{a=1}^d\,\cos q_a \in [-1,1],\label{gamma_q}
\ee
with $q_a$ the components of the wavevector $\mathbf{q}$. At the transition $a=b$ and the spectrum becomes 
gapless at ${\mathbf{q}}=0$. In principle, at the same level of approximation 
one should also take into account the coupling between the spin-waves of the Ising model 
and the conduction electrons via the hopping term in \eqn{H-Ising-Hubbard-2}. We just mention that, 
deep in the insulating side, where $\omega_{\mathbf{q}} \sim u/2 \gg 1$, one can integrate out 
the acoustic modes and obtain the antiferromagnetic Heisenberg model known to be the large $U$ 
limit of the half-filled Hubbard model. A thorough analysis of the role of quantum fluctuations 
at equilibrium has been presented in Ref.~\onlinecite{Z2-2} in connection with the $Z_2$-slave-spin 
theory for correlated fermions, to which we refer for further details. 
In what follows, we shall instead discuss a way to add quantum fluctuations 
in an out-of-equilibrium situation.  

\subsection{Recovering the Gutzwiller approximation out-of-equilibrium}
\label{Recovering the Gutzwiller approximation out-of-equilibrium}

Because the two models can be mapped onto each other, a quantum quench in the Hubbard model 
is equivalent to suddenly change the transverse field in the Ising-like model \eqn{H-Ising-Hubbard-2} 
at particle-hole symmetry and in the limit of infinite coordination lattices. We shall keep 
assuming a factorized time-dependent trial wavefunction $|\Phi_\sigma(t)\rangle\,|\Phi_c(t)\rangle$, each 
component $|\Phi_\sigma(t)\rangle$ and $|\Phi_c(t)\rangle$ being translationally 
invariant. The electron wavefunction will evolve under the action of a time-dependent hopping, which 
is however still translationally invariant. Hence, if $|\Phi_c(t=0)\rangle$ is eigenstate of the 
hopping at $t<0$, in particular its ground state state, it will stay unchanged under the time evolution. 
Therefore we shall only focus on the evolution of the Ising component. Its Hamiltonian at positive times 
and in units of $U_c$ is 
\be
H = -\frac{u_f}{4}\sum_\bR\,\Big(1-\sigma^z_\bR\Big) - 
\frac{1}{8}\,\frac{2}{z}\sum_{<\bR\,\bRp>}\,\sigma^x_\bR\,\sigma^x_\bRp,\label{H-Ising-out}
\ee
and we assume that at time $t=0$ $|\Phi_\sigma(t=0)\rangle$ is the approximate ground state defined 
in the previous section \ref{Recovering the Gutzwiller approximation at equilibrium} 
for a different transverse field $u_i$. The time-evolution is thus described by the  
Schr{\oe}dinger equation 
\be
i\partial_t\,|\Phi_\sigma(t)\rangle = H\,|\Phi_\sigma(t)\rangle.\label{Schroe-1}
\ee
We assume 
\be
|\Phi_\sigma(t)\rangle = \mathcal{U}(t)\,|\Phi_0(t)\rangle,
\ee
where now 
\be
\mathcal{U}(t) = \exp\left(i\frac{\alpha(t)}{2}\sum_i\,\sigma^x_\bR\right)\,
\exp\left(i\frac{\beta(t)}{2}\sum_\bR\,\sigma^y_\bR\right).\label{trial-Ising-out}
\ee   
It follows that $|\Phi_0\rangle$ must satisfy the equation of motion
\be
i\partial_t\,|\Phi_0(t)\rangle = H_*(t)\,|\Phi_0(t)\rangle,\label{Schroe-2}
\ee
where, apart from constants,  
\bea
H_*(t) &=& -i\,\mathcal{U}(t)^\dagger\,\dot{\mathcal{U}}(t) + \mathcal{U}(t)^\dagger\,H\,\mathcal{U}(t)
\label{H*-out}\\
&=& \sum_\bR\,\Bigg[\fract{\dot{\alpha}}{2}\,\cos\beta\,\sigma^x_\bR 
- \fract{\dot{\alpha}}{2}\,\sin\beta\,\sigma^z_\bR + \fract{\dot{\beta}}{2}\,\sigma^y_\bR\nonumber\\
&& - \fract{u_f}{4}\bigg(\cos\alpha\,\cos\beta\,\sigma^z_\bR + \cos\alpha\,\sin\beta\,\sigma^x_\bR
\nonumber \\
&&~~~~~~~~~~~~~~~~~~~~- \sin\alpha\,\sigma^y_\bR\bigg)\Bigg]\nonumber\\
&& -\frac{1}{8}\,\frac{2}{z}\sum_{<\bR,\bRp>}\,\Big(\cos\beta\,\sigma^x_\bR - \sin\beta\,\sigma^z_\bR\Big)\nonumber\\
&& \phantom{-\frac{1}{8}\,\frac{2}{z}\sum_{<i,j>}}\;\;\;
\Big(\cos\beta\,\sigma^x_\bRp - \sin\beta\,\sigma^z_\bRp\Big).\nonumber
\eea
In the same spirit of the spin-wave approximation above, we shall assume that $|\Phi_0(t)\rangle$ 
is at any time close to a fully polarized state along $x$, so that we can safely use the 
approximate expressions \eqn{sigma^x}--\eqn{sigma^z} for the spin operators. Just like before, 
we fix $\alpha(t)$ and $\beta(t)$ in such a way that all linear terms in $x_\bR$ and $p_\bR$ vanish 
and find the following set of equations
\bea
\dot{\beta} &=& -\frac{u_f}{2}\,\sin\alpha,\label{dot-beta}\\
\dot{\alpha} &=& \frac{1}{2}\,\cos\beta - \frac{u_f}{2}\,\cos\alpha\,\cot\beta.\label{dot-alpha}
\eea 
These equations have to be solved starting from the initial condition appropriate to the 
approximate ground state with transverse field $u_i$, i.e. $\alpha(0)=0$ and $\sin\beta(0)=u_i$ 
if $u_i<1$ otherwise $\sin\beta(0)=1$, see Eq.~\eqn{GA-sol-1}. In addition, as noticed before, 
the equations admit a constant of motion, which can be regarded as the classical energy, 
\[
E = -\frac{u_f}{4}\,\cos\alpha\,\sin\beta - \frac{1}{8}\,\cos^2\beta.
\]
One can readily recognize that 
the dynamical system \eqn{dot-beta}--\eqn{dot-alpha} is equivalent to that one we previously obtained 
within the time-dependent Gutzwiller approximation. However, as we are going to see in the next section,
this alternative formulation however allows us to access quantum fluctuations, assuming they are small.

\subsection{Quantum Fluctuations beyond mean field dynamics}\label{sect:fluct}

The time dependent hamiltonian $H_{\star}(t)$ we have obtained in the previous section, Eq~\eqn{H*-out}, 
accounts in principle for quantum fluctuation effects. A simple way to proceed is to fix the parameters 
$\alpha(t)$ and $\beta(t)$ in such a way Eqs.~\eqn{dot-beta} and \eqn{dot-alpha} are satisfied, 
and expand the hamiltonian up to second order 
in $x_\bR$ and $p_\bR$. The result has no more linear terms and simply describes
coupled harmonic oscillators with time-dependent parameters.
\bea
H_*(t) &\simeq& \fract{u_f\cos\alpha(t)}{4\sin\beta(t)}\,\sum_\bR\Big(x_\bR^2+p_\bR^2\Big) \nonumber\\
&& - \fract{\sin^2\beta(t)}{4}\,\frac{2}{z}\,\sum_{<\bR,\bRp>}\,x_\bR\,x_\bRp,\label{H*-out-bis}\,
\eea
We note that such a treatment, similar to what we have done in equilibrium, 
is equivalent to include gaussian fluctuations without renormalizing 
the transition point. In other words, we are studying 
the effect of quantum fluctuations around the semiclassical 
trajectory without allowing any \emph{feedback} 
of these on the latter, which could be dangerous, as we shall see.  
We shall analyze the time dependent problem (\ref{H*-out-bis}) 
separately in the two different cases of 
quenching from the correlated metal or from the Mott insulator, starting from the latter that is simpler. 

\subsubsection{Quenching from the Mott insulator}

In this case $u_i>1$ and the initial values of the Euler angles are $\alpha(0)=0$ and 
$\sin\beta(0)=1$. It follows from Eqs.~\eqn{dot-alpha} and \eqn{dot-beta} that these angles 
will not evolve in time so that $H_*$ in \eqn{H*-out-bis} does not depend on time and coincides with 
\eqn{H_*-metal-insulator} for $a=u_f/2$ and $b=1/2$.  This Hamiltonian is well defined provided $u_f>1$, 
which simply reflects that our assumption of weak quantum fluctuations loses its validity if the quench 
is too big. Therefore we shall assume $u_f>1$, namely a quench withing the Mott insulator domain. 

Initially the system is described by the Hamiltonian \eqn{H_*-metal-insulator} with $a=u_i/2$. 
We assume that the initial state is the ground state of such a Hamiltonian. At times $t>0$, this 
state is let evolve with the same Hamiltonian, but now with $a=u_f/2$. 
This problem can be readily solved, being equivalent to starting from the ground state of a 
harmonic oscillator and evolving it with a Hamiltonian having different mass and spring constant.   
We find that the time-dependent average value of the double occupancy is 
\bea
D(t) &=& \frac{1}{16V}\sum_{\bq}\,\Bigg[
\left( K_{i\bq} + \fract{1}{K_{i\bq}}
+ \fract{K_{f\bq}^2}{K_{i\bq}}+ \fract{K_{i\bq}}{K_{f\bq}^2} -4 \right)\nonumber\\
&+& \left( K_{i\bq} + \fract{1}{K_{i\bq}}
- \fract{K_{f\bq}^2}{K_{i\bq}} - \fract{K_{i\bq}}{K_{f\bq}^2}\right)\,\cos 2\omega_\bq t\Bigg], 
\label{D(t)-fluct}
\eea
where 
\[
\omega_\bq = \frac{1}{2}\sqrt{u_f\Big(u_f-\gamma_q\Big)},
\]
see \eqn{omega_q} and \eqn{gamma_q}, while 
\[
K_{i\bq}^2 = \fract{u_i}{u_i-\gamma_\bq},\qquad K_{f\bq}^2 = \fract{u_f}{u_f-\gamma_\bq},
\]
are the parameters of the canonical transformation to find the normal modes of the
initial and final Hamiltonians, i.e. $x\to \sqrt{K}\,x$ and $p\to p/\sqrt{K}$.
Seemingly, the hopping renormalization factor $Z(t)$ turns out to be
\bea
Z(t) &=& \langle \sigma^x_i\sigma^x_j\rangle \nonumber \\
&=&  \frac{1}{2V}\sum_{\bq}\,\gamma_\bq\Bigg[ 
\left(K_{i\bq}+\fract{K_{f\bq}^2}{K_{i\bq}}\right) \nonumber \\
&& + \left(K_{i\bq}-\fract{K_{f\bq}^2}{K_{i\bq}}\right)\,\cos 2\omega_\bq t\Bigg].\label{Z(t)-fluct}
\eea
We note that the sum of the oscillatory terms in \eqn{D(t)-fluct} and \eqn{Z(t)-fluct}  
vanishes for $t\to\infty$, unless $u_f\to\infty$, 
so that asymptotically $D(t\to\infty)$ and $Z(t\to\infty)$ approach
values that do not corresponds either to the initial ones nor to the equilibrium values for $u=u_f$.    

We remark that the above time evolution derives just by the quantum fluctuations. Should we neglect 
these latter, we would not find any dynamics for these quantities. 

\subsubsection{Quenching from the metal}

We now consider the case in which $u_i<1$ so that initially $\alpha(0)=0$ and $\sin\beta(0)=u_i$. 
With such initial values, the time evolution controlled by \eqn{dot-alpha} and \eqn{dot-beta} 
is non trivial, unlike the previous example of a Mott insulating initial state. 
As we mentioned before, the Hamiltonian $H_*$ describes coupled harmonic oscillators with time dependent parameters.
The time dependent frequency of these oscillations reads
\be
\omega^2_{\mathbf{q}}(t)=\frac{u_f\,\cos\alpha(t)}{\sin^2\beta(t)}-\gamma_{\mathbf{q}}\,\sin^2\beta(t)\,,
\ee
with $\gamma_{\mathbf{q}}$ defined in Eq.~(\ref{gamma_q}). Since the minimum frequency is obtained for $\mathbf{q}=0$
we immediately realize that in order to have stable fluctuations the condition  
$u_f\,cos\alpha(t)> cos^3\beta(t)$ has to hold. 

\begin{figure}[t]
\begin{center}
\epsfig{figure=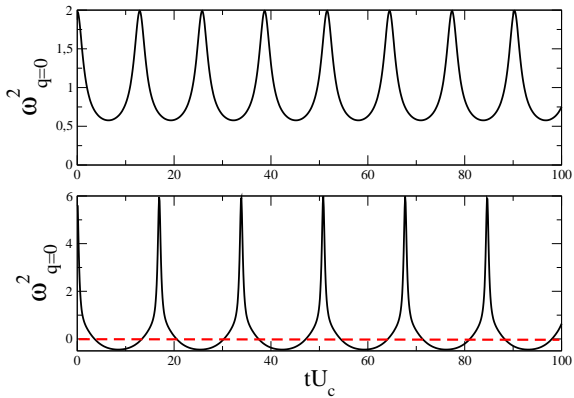,scale=0.3}
\caption{Behavior of the frequency $\omega^2_{\mathbf{q=0}}$ as a function of time for $u_i=0.1$ and $u_f=0.2$ (top panel)
and $u_f=0.6$ (bottom panel). We see that for suitable values of $u_f$ the frequency can become negative 
for some time intervals.}
\label{fig:fig10}
\end{center}
\end{figure}

In figure \ref{fig:fig10} we plot the behavior of $\omega^2_{\mathbf{q}=0}(t)$ as obtained from 
the semiclassical dynamics. We notice that for suitable values of $u_f$ it exist multiple time intervals at 
which $\omega^2_{\mathbf{q}=0}(t)<0$ and fluctuations become unstable.
\begin{figure}[t]
\begin{center}
\epsfig{figure=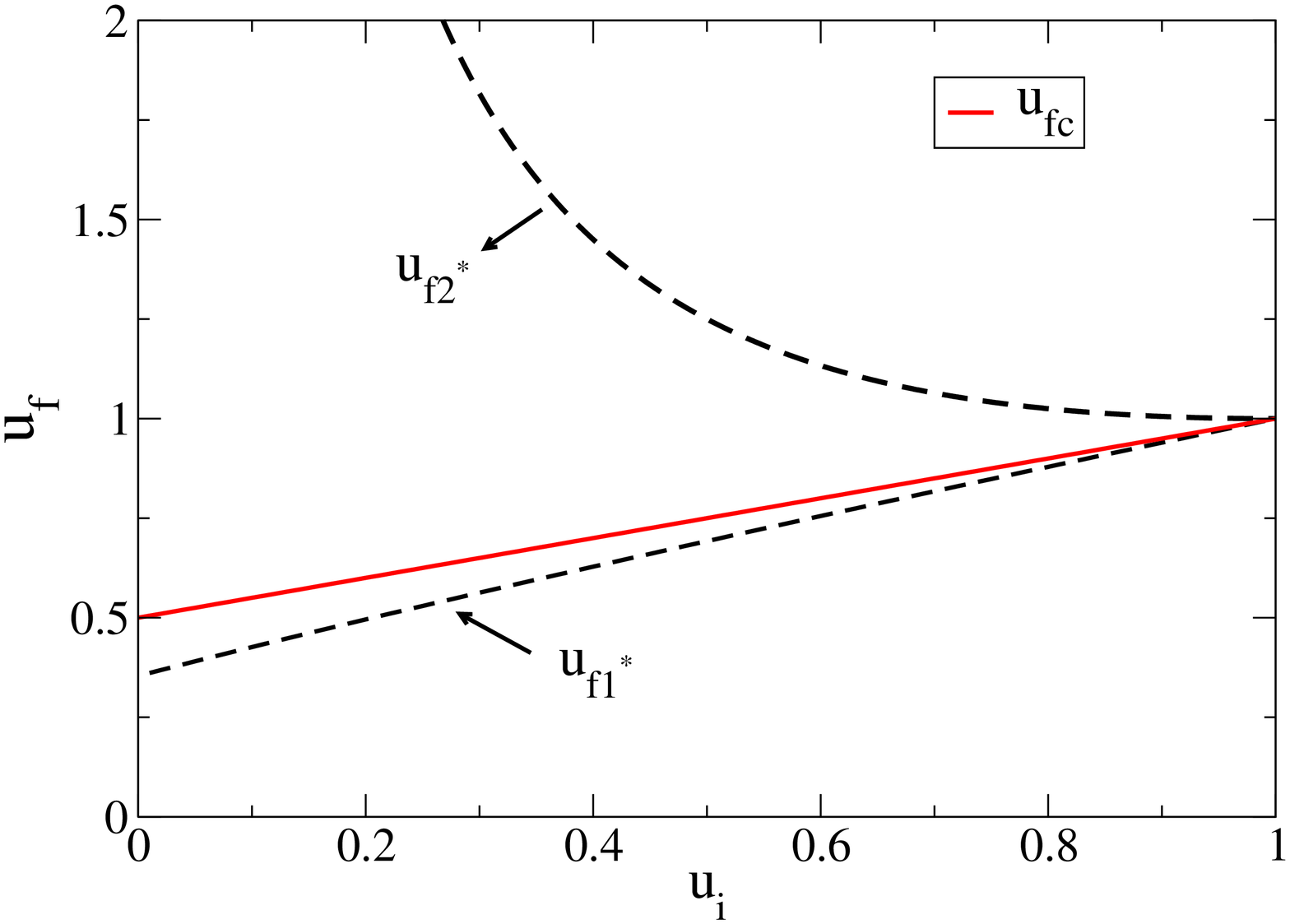,scale=0.3}
\caption{Behavior of the instability lines $u_{f1,2}^{\star}$ defined in the main text, as a function of $0<u_i<1$.
We see that these lines bound a region of the phase diagram around the mean field critical line $u_{fc}$ where
fluctuations grows exponentially in time. This region shrinks upon approaching $u_i\rightarrow1$ while becoming
wider and wider in the opposite regime of
quenches from a non interacting Fermi system.}
\label{fig:fig11}
\end{center}
\end{figure}
In particular, by looking at the mean field dynamics, it is easy to realize that there is a whole region of quenches, 
just around the dynamical transition, for which an instability in the fluctuation spectrum may occur. This region of 
unstable modes is bounded by two lines $u_{f1}^{\star}$, $u_{f2}^{\star}$ whose behavior is plotted in figure 
\ref{fig:fig11}. The line $u_{f2}^{\star}$ can be obtained analytically by simple means and reads
\be\label{eqn:uf2}
u_{f2}^{\star}= \fract{1+u_i^2}{2u_i}\,.
\ee
As a result of this analyis we conclude that for quenches \emph{below} and \emph{above} these instability lines
we can use the spin wave approximation to compute corrections to quantum dynamics,
since all $\mathbf{q}$-modes are stable. As opposite for quenches around the critical mean field line  part of 
the spectrum becomes unstable. Conversely, we previously 
found that the same method is, at least, well defined when quenching from the Mott insulator down 
to the Mott transition.  We believe that this difference is not accidental and that the dynamics of quantum fluctuations 
quenching from the metallic side is poorly described by the Hamiltonian \eqn{H*-out-bis}. 
The metallic phase corresponds in our language to the ordered phase of the Ising model, where a finite 
order parameter is spontaneously generated. 
The equations of motion \eqn{dot-beta} and \eqn{dot-alpha} describe the dynamics 
of the {\sl condensate} alone. The approach in 
section~\ref{Recovering the Gutzwiller approximation out-of-equilibrium} implicitly assumes quantum 
fluctuations that follow adiabatically the evolution of the condensate. However,  
the quantum fluctuations must in turn affect the evolution of the condensate, 
a feedback that is absent in the above scheme and explains why the latter fails if the quench is big enough. 
Anyway, the fact that the Hamiltonian \eqn{H*-out-bis} become unstable before the dynamical critical point is 
encountered suggests that the effect of quantum fluctuations grows and it is not unlikely to modify 
substantially the dynamics.

\section{Conclusions}\label{sect:conclusion}

We have introduced a variational approach to strongly correlated electrons out of equilibrium.
The idea is to give an ansatz on the time dependent many-body wave function and to obtain dynamical equations
 for the parameters by imposing a saddle point on the real-time action. While this strategy is widely used 
for non interacting fermionic systems, in the spirit of time dependent Hartree-Fock, its extension to strongly
 correlated electrons represents a novelty with many possibilities for further developments. Applications
of this method can range from dynamics in closed quantum systems to non equilibrium transport in correlated 
quantum dots, for which a related variational approach for the steady has been recently proposed~\cite{Lanata_PRB10}.

 In this paper we have applied this variational scheme to the single band Hubbard model using a proper generalization 
of the Gutzwiller wavefunction. It is worth mentioning, however, that the method is general and can be applied also
 to other correlated wavefunctions, as long as a suitable numerical or analytical approach is available to calculate 
the \emph{variational energy} controlling the classical dynamics of the variational parameters. 
As a first application we have studied the dynamics of the Hubbard model after a quantum quench of the interaction. 
This is an interesting open problem for which results have been obtained only very recently using sophisticated non 
equilibrium many body techniques. Remarkably, although extremely
 simple, our approach seems to capture many non trivial effects of the problem and shows a good overall agreement with 
the picture provided by DMFT. From this perspective it can be seen as a simple and intuitive mean field theory for quench
 dynamics in interacting Fermi systems.

\subsection*{Acknowledgment}

We would like to acknowledge interesting discussions with G. Biroli, M. Capone, C. Castellani, E. Demler, A. Georges, 
D. Huse, S. Kehrein, C. Kollath, A. Mitra, D. Pekker and E. Tosatti. This work has been supported by Italian 
Ministry of University and Research, through a PRIN-COFIN award.
\appendix

\section{Details on the Gutzwiller calculations at finite doping}\label{app:doping}

In this appendix we describe in some detail the analyis of the mean field dynamics at finite doping. We
start from the equation (\ref{eqn:effective1}) where the phase $\phi$ is expressed in terms of $D$
using energy conservation 
\be
cos^2\phi = \frac{E_0-UD-2\bar{\eps}\left(n-2D\right)\left(\sqrt{D+\delta}-\sqrt{D}\right)^2}
{8\bar{\eps}\left(n-2D\right)\sqrt{D\left(D+\delta\right)}}\,.
\ee
This result can be inserted into the equation for $D(t)$, which reads after simple differentiation
\be
\frac{dD}{dt} = -8\bar{\eps}\left(n-2D\right)\sqrt{D\left(D+\delta\right)}\,sin\phi\,cos\phi\,.
\ee
After some simple algebra we end up with a differential equation for the time-dependent double occupation 
whose general structure is
\be\label{eqn:gen_dyn_docc_app}
\frac{dD}{dt} = \pm\sqrt{\Gamma(D)}\,,
\ee
where $\Gamma(D)$ can be thought as an effective potential controlling the dynamics of $D(t)$. 
Its explicit expression reads $\Gamma\left(D\right)=\Gamma_{+}\left(D\right)\Gamma_{-}\left(D\right)$ where
\be\label{eqn:gamma_pm}
\Gamma_{\pm}\left(D\right) = \pm\left[
E_0-UD-2\bar{\eps}\left(n-2D\right)\left(\sqrt{D+\delta}\pm\sqrt{D}\right)^2\right]\,.
\ee
After some lengthy but straightforward calculations it is possible to bring the function $\Gamma\left(D\right)$ 
to a polinomial form, namely to
\be\label{eqn:gamma_poli}
\Gamma\left(D\right) = \gamma_{3}\,D^3+\gamma_{2}\,D^2+\gamma_{1}\,D+\gamma_{0}\,,
\ee
where $\gamma_a$'s are coefficients depending on the initial $U_i$ and final $U_f$ interactions as well as on the doping
$\delta$. We first notice that for $\delta=0$ the expression for $\Gamma$ simplifies to read
\be
\Gamma_{\delta=0}\left(D\right) = \left(u_fD-E_0\right)\left(E_0-u_fD+2D\left(1/2-D\right)\right)\,,
\ee
where the initial energy $E_0$ reads as in Eq.~(\ref{eqn:initial_energy}). 
It is easy to very that the effective potential has three roots $D_i$, $D_{\pm}$,
the former corresponding to the \emph{equilibrium} Gutzwiller solution at $T=0$, $D_i=\left(1-u_i\right)/4$ 
while the latters given respectively by
$$
D_{+}=
\left\{
\begin{array}{ll}
u_f < u_{fc} & \frac{u_{fc}-u_f}{2}\,,\\
u_f > u_{fc} & D_i\left(1-\frac{u_{fc}}{u_f}\right).\\
\end{array}
\right.
$$
and
$$
D_{-}=
\left\{
\begin{array}{ll}
u_f < u_{fc} & D_i\left(1-\frac{u_{fc}}{u_f}\right)\,,\\
u_f > u_{fc} & \frac{u_{fc}-u_f}{2}\,.\\
\end{array}
\right.
$$
In the doped case we cannot obtain expressions as simple. However we notice that $\Gamma_+\left(D_i\right)=0$,
since by construction
\be\label{eqn:E_0_doping}
E_0 =u_f\,D_i+2\bar{\eps}\left(n-2D_i\right)\left(\sqrt{D_i+\delta}+\sqrt{D_i}\right)^2\,.
\ee
As a consequence we can write the effective potential as
\be
\Gamma\left(D\right)\equiv\left(D-D_i\right)\Phi\left(D\right)\,,
\ee
with $\Phi(D)$ that can be formally written as
\be
\Phi\left(D\right) = \gamma_3\,D^2+\left(\gamma_2+D_0\gamma_3\right)D+\left(\gamma_1+D_0\gamma_2+D_0^2\gamma_3\right)
\ee
once the definition of the effective potential as a polynomial in $D$, Eq~(\ref{eqn:gamma_poli}), is considered.
From this result we obtain for the other two inversion points $D_{\pm}$ the following result
\be\label{eqn:D_pm_dop}
D_{\mp} = \frac{\left(\gamma_2+D_i\gamma_3\right)\mp\sqrt{\Delta}}{4u_f}\,,
\ee
with $\Delta=\left(\gamma_2+D_i\gamma_3\right)^2-4\gamma_3\left(\gamma_1+D_i\gamma_2+D_i^2\gamma_3\right)$. 
The explicit expression for the coefficients $\gamma_a$ can be easily found after some simple but lengthy algebra.
These read
$$
\left\{
\begin{array}{lll}
\gamma_3 & = & -2u_f\\
\gamma_2 & = & -u_f^2 +2E_0+u_f\left(1-2\delta\right)-\delta^2/4\\
\gamma_1 & = & 2u_fE_0 +\frac{n\delta^2}{4} + \frac{u_fn\delta}{2}-E_0\left(1-2\delta\right)
\end{array}
\right.
$$
with $E_0$ given by Eq.~(\ref{eqn:E_0_doping}). 
It is interesting to note that all the dependence from the initial interaction $u_i$ is hidden into the 
Gutzwiller equilibrium solution $D_i$.
The qualitative analysis can proceed along the same lines as in the previous section, 
the only difference being that $D_i$ is not known analytically. 
By solving the equilibrium Gutzwiller problem at finite doping (see appendix) 
we can easily obtain $D_i$, hence $D_{\mp}$ through Eq~(\ref{eqn:D_pm_dop}).
 When inserted back into the previous results for $\m{T}$ and $\m{A}$ we find further
evidence that no singularity emerges for any finite $\delta$ in those quantities,
 which nevertheless features some signature of the zero doping criticality. 
In particular both $\m{T}$ and $\m{A}$ are smooth functions displayng a sharp peak around $u_{fc}$.

\bibliographystyle{apsrev}

\end{document}